\documentclass[10pt]{article}
\usepackage{amssymb,amsmath,fullpage,times,epsf}
\usepackage{mathrsfs}
\usepackage{amssymb}
\usepackage{graphics}
\usepackage{graphicx}
\usepackage{subfigure}
\usepackage{float}
\usepackage{bm}
\usepackage{amsfonts}
\usepackage{color}

\allowdisplaybreaks
\def\beginproof{\par\noindent\textbf{Proof.}~~}
\def\endproof{\ \vbox{\hrule\hbox{\vrule height1.0ex\hskip1.0ex\vrule}\hrule }\par\medskip}

\newtheorem{theorem}{Theorem}[section]

\newtheorem{proposition}[theorem]{Proposition}

\def\lam{\lambda}

\def\ra{\rightarrow}

\def\lim{\mathrm{lim}}
\def\sech{\mathrm{sech}}
\def\tanh{\mathrm{tanh}}

\def\dn{\mathrm{dn}}
\def\cn{\mathrm{cn}}
\def\sn{\mathrm{sn}}

\def\max{\mathrm{max}}
\def\PT{$\mathcal{P}\mathcal{T}$}

\def\eref#1{(\ref{#1})}

\usepackage[square,numbers,sort&compress]{natbib}
\def\Im{\mathrm{Im}}

\begin{document}
\date{}

\title{General stationary solutions of the nonlocal nonlinear Schr\"{o}dinger equation and their relevance to the \PT-symmetric system}
\author{Tao Xu$^{1,2,}$\thanks{Corresponding author, e-mail: xutao@cup.edu.cn}\,, Yang Chen$^{2}$, Min Li$^{3,}$\thanks{Corresponding author, e-mail: ml85@ncepu.edu.cn}\,, and De-Xin Meng$^{2}$ \\
{\em 1. State Key Laboratory of Heavy Oil Processing,}\\
{\em China University of Petroleum, Beijing 102249, China }\\
{\em 2. College of Science, China University of Petroleum, Beijing 102249, China}
\\{\em 3.  Department of Mathematics and Physics, }\\
{\em  North China Electric Power University, Beijing 102206, China}}
\maketitle

\begin{abstract}
With the stationary solution assumption, we establish the connection between the nonlocal nonlinear Schr\"{o}dinger (NNLS) equation and an elliptic equation.  Then, we obtain the general stationary solutions and discuss the relevance of their smoothness and boundedness to some integral constants. Those solutions, which cover the known results in the literature, include the unbounded Jacobi elliptic-function and hyperbolic-function solutions, the bounded sn-, cn- and dn-function solutions, as well as the hyperbolic soliton solutions. By the imaginary translation transformation of the NNLS equation, we also derive the complex-amplitude stationary solutions, in which all the bounded cases obey either the \PT- or anti-\PT-symmetric relation. In particular, the complex tanh-function solution can exhibit no spatial localization in addition to the dark and anti-dark soliton profiles, which is sharp contrast with the common dark soliton. Considering the physical relevance to \PT-symmetric system, we show that the complex-amplitude stationary solutions can yield a wide class of complex and time-independent \PT-symmetric potentials, and the symmetry breaking does not occur in the \PT-symmetric linear system with the associated potentials.


\vspace*{4mm}

\noindent{Keywords: Nonlocal nonlinear Schr\"{o}dinger equation; Jacobi elliptic-function solutions; Hyperbolic soliton solutions; Parity-time symmetry}

\end{abstract}

\newpage

\section{Introduction}
Recently, there has been a growing interest in the  nonlocal integrable nonlinear partial differential
equations (NPDEs)  in mathematical physics and soliton theory. In 2013, Ablowitz and Musslimani first proposed the following nonlocal nonlinear Schr\"{o}dinger (NNLS) equation~\cite{Ablowitz1}:
\begin{align}
i \,U_{t}(x,t)+U_{xx}(x,t)+\sigma\,U(x,t)^2\hat{U}(x,t)=0,\label{NNLS}
\end{align}
where $U$ is a complex-valued function of  $x$ and $t$, $\sigma=\pm1$ signifies the focusing $(+)$ and defocusing $(-)$  nonlinearity, and the $hat$ denotes the combined operation of
complex conjugate and space reversal, i.e., $\hat{U}=U^*(-x,t)$.  
The nonlinear coupling between $U(x,t)$ and $U^*(-x,t)$  in Eq.~\eref{NNLS} reflects the parity-mirror nonlocality,
which is contrast with the standard (local) nonlinear Schr\"{o}dinger (NLS) equation
\begin{align}
i \,U_{t}(x,t)+U_{xx}(x,t)+\sigma\,|U(x,t)|^2U(x,t)=0.   \label{NLS}
\end{align}
Remarkably, Eq.~\eref{NNLS} can arise from a complex reverse-space reduction of the $2\times 2$ Ablowitz-Kaup-Newell-Segur scattering problem, and thus it is a completely integrable model~\cite{Ablowitz1}. This also makes researchers realize that the reverse-space, reverse-time and reverse-space-time nonlocal reductions (which have been overlooked before) may widely exist in the known linear scattering problems, like the Ablowitz-Kaup-Newell-Segur~\cite{AKNS}, Kaup-Newell~\cite{KN} and Wadati-Konno-Ichikawa~\cite{WKI} schemes.  Soon thereafter, a number of nonlocal integrable NPDEs have been identified in both one and two space dimensions as well as in discrete settings~\cite{Ablowitz2,Ablowitz3,Ablowitz4,Ablowitz6,Ablowitz7,Fokas,vector,
Sinha,DNLS,NKN,mKdV,NNWave,SS,AB-KdV,Lou2,Tang}.

In the past few years, the mathematical properties of Eq.~\eref{NNLS} have been intensively studied from different aspects: the inverse scattering transform schemes for the initial value problems  with zero and nonzero boundary conditions~\cite{Ablowitz1,Ablowitz3,Ablowitz5}, hierarchy Hamiltonian structures for the NNLS equations~\cite{Gerdjikov},
long-time asymptotic behavior with  decaying  boundary conditions~\cite{Rybalkoy},
equivalent transformation between the NLS and NNLS equations~\cite{YJK1},  etc.
Also, various analytical methods have
been used to derive wide classes of explicit solutions for both the $\sigma=1$ and $\sigma=-1$ cases of Eq.~\eref{NNLS}~\cite{Ablowitz1,Ablowitz3,Ablowitz5,Sarma,LiXu,LiXu1,
HJS,Gupta,YJK2,YJK3,YZY,Wen,Gurses,Khare,ZDJ,FBF,LML,Michor,YChen,LiXu3,Santini}.
In contrast with the local NLS equation, the focusing case of Eq.~\eref{NNLS} possesses the bright-soliton, dark-soliton, rogue-wave and breather solutions, simultaneously~\cite{Ablowitz1,Sarma,Khare,Gupta,YJK2,YJK3,Santini}. Those solutions
in general develop the collapsing singularities in finite time, and they are bounded only for some particular parametric choices~\cite{Ablowitz1,Ablowitz3,YJK2,YJK3}. For the defocusing case, Eq.~\eref{NNLS} admits the exponential soliton solutions, rational soliton solutions, exponential-and-rational soliton solutions on the plane-wave background~\cite{LiXu,LiXu1,Wen,HJS,Ablowitz5,FBF}. However, no single-soliton behavior was found in such three types of soliton solutions although they can display a rich variety of elastic interactions among dark and anti-dark solitons~\cite{LiXu,LiXu1,LiXu3}.

Meanwhile, in order to show the physical relevance of Eq.~\eref{NNLS}, Ref.~\cite{Magstr} found that it is linked to an unconventional coupled Landau-Lifshitz system in magnetics through the gauge transformation, Ref.~\cite{AMJPA} reported that it can be derived as the quasi-monochromatic complex reductions of the cubic nonlinear Klein-Gordon, Korteweg-de Vries and water wave equations. In addition, several efforts have been made towards the physical realization of the parity-mirror nonlocality since it is quite different from the usual nonlocality of the integral type. It was suggested that the parity-mirror nonlocal coupling can be implemented in a nonlinear string where each particle is simultaneously coupled with nearest neighbors and its mirror particle~\cite{Gadzhimuradov2}, in the electrical transmission line with nonlocal nonlinear elements~\cite{Gadzhimuradov2}, and in the coupled waveguides with some parity-symmetry constraint between the two components~\cite{YJK4}.

Eq.~\eref{NNLS} is ususally said to be parity-time (\PT) symmetric since it is invariant under the combined action of parity operator $\mathcal{P}$ ($x \ra -x$) and time-reversal operator $\mathcal{T}$ ($t \ra -t$, $i \ra -i$).
Formally, one can view Eq.~\eref{NNLS} as the \PT-symmetric linear Schr\"{o}dinger (PTLS) equation:
\begin{align}
i\,U_t(x,t) + U_{xx}(x,t) + V(x,t)U(x,t) =0 \,\,\text{with}\,\, V(x,t)=\sigma\,U(x,t)\hat{U}(x,t),  \label{PTLS}
\end{align}
where $V(x,t)$ is the self-induced \PT-symmetric potential which naturally satisfies $V(x,t)=V^{*}(-x,t)$. Thus, Eq.~\eref{NNLS} may have potential applications in the \PT-symmetric quantum mechanics, optics, and other related areas~\cite{PTPhys}. Within this regard, Ref.~\cite{HJS} discussed the similarity between the time-dependent \PT-symmetric potential generated from an exact rational solution and the gain/loss distribution in a \PT-symmetric optical system; Ref.~\cite{Yan2} studied the soliton dynamics in the NNLS equation with several types of \PT-symmetric external potentials.

We notice that the stationary solutions in terms of the Jacobi elliptic functions and  hyperbolic functions of Eq.~\eref{NNLS} were reported in Refs.~\cite{Sarma,Khare}, but all of them are either even- or odd-symmetric with respect to $x$. As a matter of fact, those symmetric stationary solutions are just some special cases when the NLS and NNLS equations are satisfied simultaneously. To be specific, the even-symmetric solutions are shared by the focusing cases of Eqs.~\eref{NNLS} and~\eref{NLS} in common, whereas the odd-symmetric solutions of the focusing Eq.~\eref{NNLS} solves the defocusing Eq.~\eref{NLS} (vice versa). The main concern of this paper is to construct the general stationary solutions which can cover the known even- and odd-symmetric cases in the literature. On the other hand, it has been shown that the soliton theory can provide some useful information for the \PT-symmetric physics~\cite{HJS,Wadati,Khare2,Zhao}. With this consideration, we will use the obtained stationary solutions to construct complex \PT-symmetric potentials and discuss whether the symmetry breaking occurs in the associated \PT-symmetric linear system.


In this work, based on the connection between Eq.~\eref{NNLS} and an elliptic equation with the stationary solution assumption, we derive the general unbounded Jacobi elliptic-function and hyperbolic-function solutions which are growing to infinity at $x\to \infty$ (or $x\to -\infty$) but decaying to zero at $x\to -\infty$ (or $x\to \infty$); and also obtain the bounded sn-, cn- and dn-function solutions as well as the hyperbolic soliton solutions, which are the same as those in Refs.~\cite{Sarma,Khare}. Meanwhile, the imaginary translation transformation is applied to those obtained solutions. As a result, we obtain many complex-amplitude stationary solutions, in which all the bounded cases obey either \PT- or anti-\PT-symmetric relation (i.e., $U=$\PT$ U$ or $U=-$\PT$U$). Of special interest, the complex tanh function solution can display both the dark- and antidark-soliton profiles or exhibit no spatial localization. Moreover, we show the physical relevance of those complex-amplitude stationary solutions by using them to construct a wide class of \PT-symmetric potentials, whose associated Hamiltonians are \PT-symmetry unbroken. Different from the studies in Refs.~\cite{HJS,Wadati,Khare2}, all the obtained  \PT-symmetric potentials are  time-independent, and for the bounded cases the solutions' amplitudes correspond to some eigenstates of the PTLS equation with proper boundary condition.

The structure of this paper is organized as follows: In Section~\ref{Sec2}, 
we establish the relationship of Eq.~\eref{NNLS} with an elliptic equation, and discuss how the smoothness and boundedness of stationary solutions is related to some integral constants in the elliptic equation. In Section~\ref{Sec3},
we derive the unbounded Jacobi elliptic-function and hyperbolic-function solutions, the bounded sn-, cn- and dn-function solutions, as well as the hyperbolic soliton solutions. Meanwhile, by the imaginary translation transformation, we obtain the complex-amplitude stationary solutions, and discuss their associated nonsingular conditions.  In Section~\ref{Sec4}, we show that the complex-amplitude solutions can be used to construct the time-independent \PT-symmetric potentials, and prove that their associated Hamiltonians remain the unbroken \PT~symmetry. In Section 5, we address the conclusions and discussions of this paper.

\section{Connection between Eq.~\eref{NNLS} and an elliptic equation}
\label{Sec2}

We assume the stationary solution of Eq.~\eref{NNLS} take the form
\begin{align}
U(x,t)=u(x)e^{i \, \mu t}, \quad \hat{U}(x,t)=\hat{u}(x)e^{-i\, \mu t}, \label{StaSol}
\end{align}
where $u(x)$ is the complex-valued amplitude, $\mu$ is an arbitrary real constant. With this stationary solution assumption, Eq.~\eref{NNLS} can be reduced to
\begin{equation}
u''-\mu u +\sigma u^2 \hat{u} =0. \label{NNLS_State}
\end{equation}
Since Eq.~\eref{NNLS_State} remains invariant when $x$ changes to $-x$ and  complex conjugate is taken,
 then $\hat{u}(x)$  also satisfies Eq.~\eref{NNLS_State},
\begin{equation}
\hat{u}''-\mu\hat{u} +\sigma\hat{u}^{2} u =0. \label{NNLS_uhat}
\end{equation}
Multiplying Eq.~\eref{NNLS_State} by $2\hat{u}'(x)$ and adding its \PT-symmetric counterpart,
then integrating the resulting equation once with respect to $x$, we have
\begin{equation}
2 u' \hat{u}' - 2\mu u\hat{u} +\sigma(u\hat{u})^{2}=C_{0}, \label{NNLS_VC01}
\end{equation}
where $C_{0}$ is a real integral constant because the left-hand side of Eq.~\eref{NNLS_VC01} is itself with the  \PT~operation.
We multiply Eqs.~\eref{NNLS_State} and~\eref{NNLS_uhat} respectively by $\hat{u}$ and $u$, and add them to Eq.~\eref{NNLS_VC01}, yielding
\begin{equation}
(u\hat{u})'' -4\mu u\hat{u}  + 3\sigma (u\hat{u})^2 = C_{0},  \label{Eq8}
\end{equation}
Again, multiplying Eq.~\eref{Eq8} by $(u\hat{u})'$  and integrate the resulting equation once with respect to $x$,
we arrive at the elliptic equation for $W=u \hat{u}$ as follows:
\begin{equation}
{W'}^{2}-4\mu W^{2}+2\sigma W^{3}=2C_{0}W+C_{1}, \label{Veq}
\end{equation}
where $C_1$ is also a real integral constant.

On the other hand, one can from Eqs.~\eref{NNLS_State} and~\eref{NNLS_uhat} obtain that $\hat{u}'' u-u''\hat{u}=(\hat{u}' u-u'\hat{u})'=0$, that is,
\begin{equation}
\hat{u}' u-u'\hat{u}=C_2, \label{C2}
\end{equation}
where $C_2$ is a real constant due to the invariance of the left-hand side with the \PT~operation.
Divided by $u^2$ or $\hat{u}^2$, Eq.~\eref{C2} becomes
\begin{equation}
\left(\frac{\hat{u}}{u}\right)' = \left(\frac{W}{u^2}\right)' = \frac{C_2}{W}\cdot\frac{W}{u^2}\quad \text{or} \quad \left(\frac{u}{\hat{u}}\right)' = \left(\frac{W}{\hat{u}^2}\right)' = \frac{-C_2}{W}\cdot\frac{W}{\hat{u}^2}, \quad (W=u\hat{u}).
\label{uuh}
\end{equation}
Note that if $W$ is solved from Eq.~\eref{Veq} and satisfies the \PT-symmetric relation $W(x)=\hat{W}(x)$, then Eq.~\eref{uuh} can be viewed as a linear equation with respect to $W/u^2$ or $W/\hat{u}^2$. Thus, we  have
\begin{equation}
u^{2}= \rho^2 W(x)e^{-\int_{x_0}^{x}\frac{ C_2}{W(s)}ds},\quad \hat{u}^{2}= \frac{W(x)}{\rho^2}e^{\int_{x_0}^{x}\frac{C_2}{W(s)}ds}. \label{ComplexVu}
\end{equation}
Because $\widehat{u^{2}}(x) = \hat{u}^{2}(x)$, $\rho$ can be determined as
\begin{align}
\rho =e^{-\frac{1}{4}\int_{-x_0}^{x_0} \frac{C_2}{W(s)}ds}.
\end{align}

Next, we check whether $u(x)$ and $\hat{u}(x)$  in Eq.~\eref{ComplexVu} satisfy Eq.~\eref{NNLS_State}.
First, taking the second-derivative of $u$ yields
\begin{align}
u''= \frac{u(C_2^2-W'^2)}{4 W^2}+\frac{u W''}{2 W}. \label{utwo}
\end{align}
Then, substituting~\eref{utwo} into~\eref{NNLS_State} and removing $W'^2$ and $W''$ by Eqs.~\eref{Eq8} and~\eref{Veq}, we obtain
\begin{align}
u''-\mu u +\sigma u^2 \hat{u} = 
\frac{\left(C_2^2-C_1\right) u}{4 W^2},  \label{verif}
\end{align}
where $W=u\hat{u}$ has been used for simplification. It follows from Eq.~\eref{verif} that $u(x)$ and $\hat{u}(x)$  in
Eq.~\eref{ComplexVu} solve Eq.~\eref{NNLS_State} if and only if $C_1 = C_2^2 \ge 0$.
Therefore, we finally arrive at the following result:

\begin{proposition} \label{proposition1}
Suppose that $W(x)$ is a \PT-symmetric solution of Eq.~\eref{Veq} with $C_1 \ge 0 $, and that the square root of $W(x)$ is a smooth function satisfying $\widehat{W^\frac{1}{2}}(x) = W^\frac{1}{2}(x)$ and $W(0)\ge 0$. Then, Eq.~\eref{NNLS_State} admits a pair of solutions:
\begin{align}
u= \rho  W^\frac{1}{2}(x)e^{-\frac{1}{2}\int^{x}_{x_0}\frac{C_2}{W(s)}ds}, \quad
\hat{u}= \frac{1}{\rho}  W^\frac{1}{2}(x)e^{\frac{1}{2}\int^x_{x_0}\frac{C_2}{W(s)}ds},
\label{ComplexVu11}
\end{align}
where $\rho =e^{-\frac{1}{4}\int_{-x_0}^{x_0} \frac{C_2}{W(s)}ds} $ and $C_2 = \pm \sqrt{C_1}$. If $C_2 \neq 0$, $u$ and $\hat{u}$ are two linearly-independent solutions; but they coalesce into one solution at $C_2 = 0$.
\end{proposition}

\textit{Remark 1.} If $W(x)$ is a real-valued  solution of Eq.~\eref{Veq} with $C_1\ge0$ and satisfies $W=\hat{W}$, it must be an even function of $x$. Due to the square-root operation in Eq.~\eref{ComplexVu11}, not all the real-valued, even-symmetric  $W(x)$ can be used to obtain the \emph{smooth} solutions $u(x)$ and $\hat{u}(x)$ for Eq.~\eref{NNLS_State}. In fact, there are two \emph{necessary} conditions to ensure the smoothness of $u(x)$ and $\hat{u}(x)$:
\begin{align}
\begin{array}l
\text{(i)} \,  W(0)=u(0)\hat{u}(0)=|u(0)|^2 \ge 0; \\[2mm]
 \text{(ii)}\, W(x)\ge 0\,\,  \text{or}\,\,  W(x)\le 0\,\,   \text{for all}\,\,  x\in  \mathbb{R}.
\end{array} \label{Cond}
\end{align}
Here, condition~(ii) results from that the sign indefiniteness of $W(x)$ may cause $u(x)$ and $\hat{u}(x)$ non-smooth at points where the sign changes. However, some exceptions still exist even if these two necessary conditions hold at the same time. For example, with $C_0=\mu ^2\,,C_1=0$ and $\sigma=-1$, Eq.~\eref{Veq} admits the following solution:
\begin{align}
W(x) = -\mu\,\tanh ^2\left(\sqrt{\frac{-\mu}{2}}\,x\right), \quad  (\mu<0), \label{ExamVx}
\end{align}
which is nonnegative for all $x\in \mathbb{R}$. But when extracting the square roots of $W(x)$ in Eq.~\eref{ExamVx}, all the smooth results violate the relation $\widehat{W^\frac{1}{2}}(x) = W^\frac{1}{2}(x)$.

\textit{Remark 2.} If $W(x) $ is a real-valued, even-symmetric solution of Eq.~\eref{Veq} with $C_1 \ge 0$ and satisfies conditions~\eref{Cond} and $\widehat{W^\frac{1}{2}}(x) = W^\frac{1}{2}(x)$, one can write  $ u(x)= W^\frac{1}{2}(x)e^{-\frac{1}{2}\int^{x}_0\frac{C_2}{W(s)}ds}$  and $ \hat{u}(x)= W^\frac{1}{2}(x)e^{\frac{1}{2}\int^{x}_0\frac{C_2}{W(s)}ds}$. Moreover, if $u(x)$ or $\hat{u}(x)$ is  bounded for all $x\in \mathbb{R}$, then $C_2=0$; in other words, $C_2=0$ is a necessary condition for $u(x)$ and $\hat{u}(x)$ to be  globally bounded on $\mathbb{R}$. The arguments are as follows: Let us assume that
\begin{align}
|u(x)|= |W^\frac{1}{2}(x)|e^{-\frac{1}{2}\int^{x}_0\frac{C_2}{W(s)}ds}\leq M, \quad  (x\in \mathbb{R}), \label{Inequa1}
\end{align}
where $M$ is a positive number. Noticing that $W(x)=W(-x)$ and $W^\frac{1}{2}(x)$ is an even- or odd-symmetric function of  $x$, then another inequality can be derived as
\begin{align}|u(-x)|= |W^\frac{1}{2}(x)|e^{-\frac{1}{2}\int^{-x}_0\frac{C_2}{W(s)}ds} = |W^\frac{1}{2}(x)|e^{\frac{1}{2}\int^{x}_0\frac{C_2}{W(s)}ds} \leq M, \quad (x\in \mathbb{R}). \label{Inequa2}
\end{align}
Combining the inequalities~\eref{Inequa1} and~\eref{Inequa2} and thanks to the sign-definiteness of $W(x)$, we have
\begin{align}
1\leq \frac{M^2}{|W(x)|} e^{-\int^{x}_0\frac{|C_2|}{|W(s)|}ds}, \quad  (x\in \mathbb{R}).    \label{Inequa3}
\end{align}
Multiplying Eq.~\eref{Inequa3} by $|C_2|$ and integrating both sides from $0$ to $x'$ gives
\begin{align}
 |C_2|x' \leq M^2\Big(1-e^{-\int_0^{x'}|\frac{C_2}{W(s)}|ds}\Big)\leq M^2.
\end{align}
However, this cannot hold true for any $x'\in \mathbb{R}$ unless $C_2=0$.

\textit{Remark 3.} With ``$hat$'' replaced by ``asterisk'', proposition~\ref{proposition1} also applies to the NLS
equation~\eref{NLS}. The difference lies in that $C_2$ must be a purely imaginary number or $0$. 
Correspondingly, $C_1$ ought to be a nonpositive number since $C_1 = C_2^2$ is still required. Accordingly,  if $W(x)$ is a real-valued, nonnegative solution of Eq.~\eref{Veq} with $C_1 \le 0$ and the smoothness of $W^{\frac{1}{2}}(x)$ is assured for $x\in \mathbb{R}$, one can obtain that \begin{align}
u=  W^\frac{1}{2}(x)e^{-\frac{1}{2}\int^{x}_{x_0}\frac{C_2}{W(s)}ds},\quad u^*=  W^\frac{1}{2}(x)e^{\frac{1}{2}\int^x_{x_0}\frac{C_2}{W(s)}ds},
\end{align}
exactly solve the equation $ u''-\mu u +\sigma |u|^2 u =0$ which is reduced from Eq.~\eref{NLS} with the stationary solution assumption. It should be mentioned that for the particular case $C_2=0$,  $u(x)$ and $\hat{u}(x)$ in Eq.~\eref{ComplexVu11} are the same and satisfy both the NLS  and NNLS equations.

\section{General elliptic-function and hyperbolic-function stationary solutions}
\label{Sec3}
With the transformation
\begin{align}
W(x)=\frac{2\mu}{3\sigma}-\frac{2}{\sigma} W_{1}(x), \label{Trans}
\end{align}
Eq.~\eref{Veq} can be changed into the standard Weierstrass elliptic equation:
\begin{equation}
W_1^{\prime2}(x)=4W^{3}_{1}(x)-g_{2}W_{1}(x)-g_{3},\label{W_KdvNom}
\end{equation}
where $g_{2}=\sigma C_{0}+\frac{4}{3}\mu^{2} $ and $g_{3}=-\left(\frac{1}{3}\sigma \mu C_{0} + \frac{1}{4}C_1 +\frac{8}{27}\mu^{3}\right)$.
It is known that Eq.~\eref{W_KdvNom} has the Jacobi elliptic-function solution~\cite{whittaker}:
\begin{align}
W_1(x)=r_{3}+(r_{2}-r_{3})\sn^2(\sqrt{r_{1}-r_{3}}\,x+  z_0, m),\quad   \left(m=\frac{r_{2}-r_{3}}{r_{1}-r_{3}}\right), \label{SNsolution}
\end{align}
where  $z_0$ is a complex number, $r_i$'s ($1\leq i \leq 3$) are the roots of the cubic equation
\begin{align}
f(r):=4r^{3}-g_{2}r-g_{3}=0,
\end{align}
with $r_1+r_2+r_3=0$. Then, inserting~\eref{SNsolution} into the transformation~\eref{Trans}, we have
\begin{align}
W(x)=2\sigma\left[\frac{\mu}{3}-r_{3}-(r_{2}-r_{3})\sn^2(\sqrt{r_{1}-r_{3}}\,x + z_0, m) \right], \quad \Big(m =\frac{r_{2}-r_{3}}{r_{1}-r_{3}} \Big), \label{Vsolution}
\end{align}
which can be substituted into Eqs.~\eref{ComplexVu11} and~\eref{StaSol} for obtaining the stationary solutions of Eq.~\eref{NNLS}.

Throughout this section, we just consider that $r_i$'s ($1\leq i \leq 3$) are real numbers so that the parameter
$m \in \mathbb{R}$, and without loss of generality assume that $r_{1}\geq r_{2}\geq r_{3}$.
Thus, we must require the modular discriminant
\begin{align}
\Delta:=g_{2}^{3}-27g_{3}^{2} = -\frac{27}{16}C^2_1 -\left(\frac{9}{2} C_0 \mu  \sigma + 4 \mu ^3 \right)C_1
+ C_0^2(\mu ^2+\sigma C_0) \ge 0.
\end{align}
This condition is satisfied if and only if $C_0$, $C_1$ and $\mu$  obey
\begin{align}
\left\{
\begin{array}l
\Xi := \sqrt{\left(3\sigma C_{0}+4\mu^{2}\right)^3} \ge 9\sigma\mu C_0 + 8\mu^3,  \\[2mm]
\max\Big\{0, -\frac{4}{27}\left(9\sigma C_{0}\mu+ 8\mu^{3} + \Xi \right)\Big\} \leq C_{1}\leq  - \frac{4}{27}\left(9\sigma C_{0}\mu + 8\mu^{3}-\Xi \right).
\end{array}
\right.  \label{Cond1}
\end{align}
On the other hand, proposition~\ref{proposition1} says that the \PT-symmetric relation $\hat{W_{1}}=W_{1}$ should hold, which implies that
\begin{align}
\sn^2(-\sqrt{r_{1}-r_{3}}\,x +z^*_0, m) = \sn^2(\sqrt{r_{1}-r_{3}}\,x + z_0, m).
\end{align}
Letting $z_0 = \gamma+ i \delta$ and according to the properties of the $\sn$-function,
we know that $\delta$ could be an arbitrary real number,
but $\gamma$ is restricted to be $ 0 $ or $K$~\footnote[2]{There is no need to consider the cases $\gamma=nK$ ($n> 1, \, n\in \mathbb{Z}$) in view of the periodicity of $\sn^2(x)$.} with $K$ being the complete elliptic integral of the first kind
\begin{align}
K=K(m)=\int_{0}^{\frac{\pi}{2}}\frac{dt}{\sqrt{1-m\sin^2t}} .  \label{ComEllInt}
\end{align}


Therefore, when $C_0$, $C_1$ and $\mu$ meet conditions~\eref{Cond1},  one can obtain the general Jacobi elliptic-function solutions and their reduced hyperbolic-function solutions for Eq.~\eref{NNLS} by substituting  Eq.~\eref{Vsolution} into Eq.~\eref{ComplexVu11} and using Eq.~\eref{StaSol}. As described in remark 1 below proposition~\ref{proposition1},  $W(x)$ should also satisfy $\widehat{W^\frac{1}{2}}(x) = W^\frac{1}{2}(x)$ and the two necessary conditions in~\eref{Cond}, so as to assure the smoothness of $U$ and $\hat{U}$. In what follows, we discuss the nontrivial smooth stationary solutions in two cases: Jacobi elliptic-function solutions ($r_1 > r_2 >r_3$) and hyperbolic-function solutions ($r_1 = r_2 >r_3$)~\footnote[3]{It is trivial for another particular case $r_1>r_2=r_3$ because one can just obtain the constant solution for $W(x)$ from Eq.~\eref{Vsolution}.}.


\subsection{Jacobi elliptic-function solutions with $\delta=0$}
\label{sec3.1}

In this subsection, we restrict $\delta=0$ and reveal all the possible Jacobi elliptic-function solutions when $r_1 > r_2 >r_3$.

First, we consider $\sigma=1, \gamma=0$ and $W(x)\ge 0$ for all $x\in \mathbb{R}$. As $0\le \sn^2(\sqrt{r_{1}-r_{3}}\,x,m)\le 1$ and $r_2 >r_3$, one immediately have $r_2 \leq \frac{1}{3}\mu $. Meanwhile, noticing that
\begin{align}
f(r)< 0\,\, \text{for}\,\,r\in(r_2, r_1),\,\,\, f(r)> 0\,\,\text{for}\,\,r\in(r_1,\infty)\,\,\text{and}\,\,
f\big(\frac{1}{3}\mu\big)=\frac{C_1}{4}\ge 0 \,\,(C_1=C^2_2), \notag
\end{align}
we then obtain the following three cases satisfying that $\widehat{W^\frac{1}{2}}(x) = W^\frac{1}{2}(x)$ and $W(x)\ge 0$ for all $x\in \mathbb{R}$.

\begin{figure}[H]
\centering
\subfigure[]{\label{Fig1a}
\includegraphics[width=2.4in]{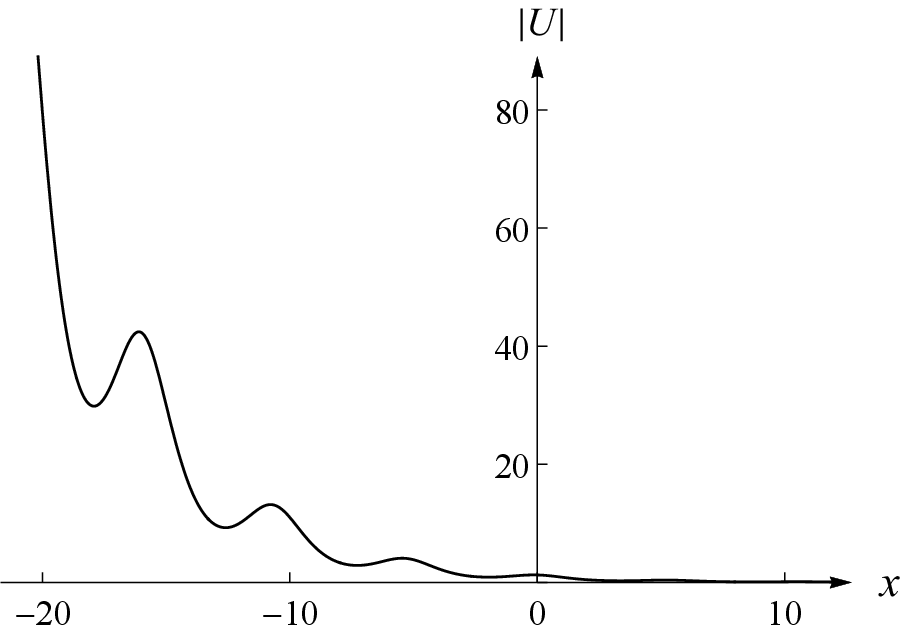}}\hspace{2.5cm}
\subfigure[]{ \label{Fig1b}
\includegraphics[width=2.4in]{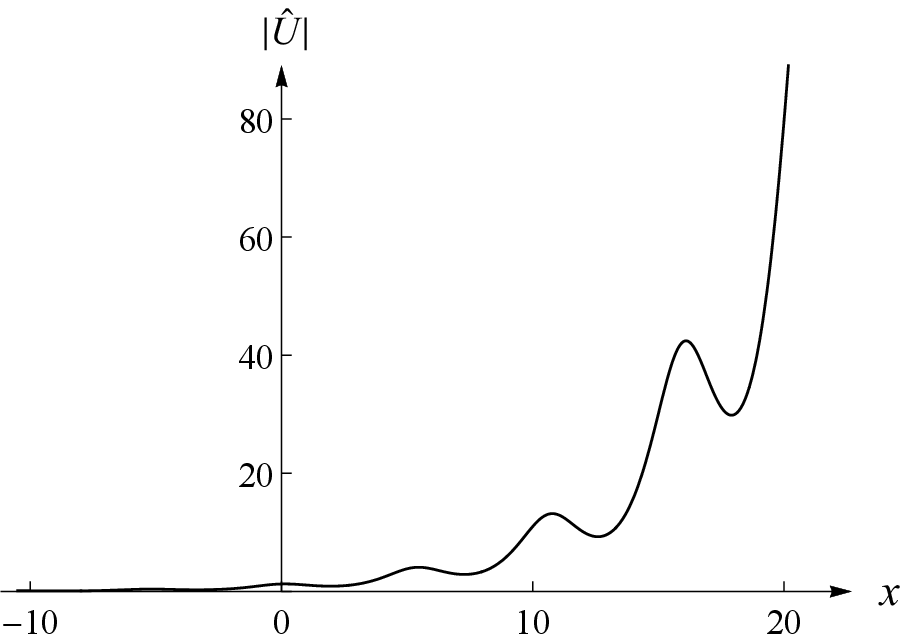}}
\caption{\small  Profiles of the amplitudes in $U$ and $\hat{U}$ via solution~\eref{ux}  with $C_0 = -\frac{2 }{3}$, $C_1=\frac{2}{27}$, $C_2= \sqrt{\frac{2}{27}}$  and $\mu=1$. } \label{Fig1}
\end{figure}

\begin{enumerate}
\item[(i)]
If $0<r_1 < \frac{1}{3}\mu$ ($\mu>0$) and $C_1>0$, we have
\begin{align}
\left\{
\begin{array}l
U_{\rm{i}}= \sqrt{\frac{2\mu}{3} - 2r_{3}-2(r_{2}-r_{3})\sn^2(\sqrt{r_{1}-r_{3}}\,x, m)}\,e^{\frac{- 3  C_2 \Pi \big(\frac{3 \left(r_2-r_3\right)}{\mu -3 r_3}; \phi(x), m\big)}{4 \left(\mu -3 r_3\right) \sqrt{r_1-r_3}} + i \, \mu t},  \\[2mm]
\hat{U}_{\rm{i}}= \sqrt{\frac{2\mu}{3} - 2r_{3}- 2(r_{2}-r_{3})\sn^2(\sqrt{r_{1}-r_{3}}\,x, m)}\,e^{\frac{ 3  C_2 \Pi \big(\frac{3 \left(r_2-r_3\right)}{\mu -3 r_3}; \phi(x), m\big) }{4\left(\mu -3 r_3\right)\sqrt{r_1-r_3}} - i \, \mu t},
\end{array}\right.\label{ux}
\end{align}
where $m =\frac{r_{2}-r_{3}}{r_{1}-r_{3}}$, $\phi(x)=\text{am}\left(\sqrt{r_1-r_3}\,x, m\right)$ is the Jacobi amplitude, $\Pi$ represents the incomplete elliptic integral of the third kind. In such generic case, the amplitude of $U$ grows to infinity at $x\to \infty$ (or $x\to -\infty$) but decays to zero at $x\to -\infty$ (or $x\to \infty$) in an exponential-and-periodical manner, so does the amplitude of $\hat{U}$. In fact, because $0<\frac{3 \left(r_2-r_3\right)}{\mu -3 r_3}<1$, $\Pi$ is a monotonically increasing function in $\mathbb{R}$, which leads to an infinite growth of the amplitudes of $U$ and $\hat{U}$. For example, with $C_0 = -\frac{2 }{3}\mu ^2$, $C_1=\frac{2}{27} \mu^3$ and $C_2=\pm \sqrt{\frac{2}{27}}\, \mu^{\frac{3}{2}}$, one can obtain $r_1=\frac{\sqrt{21}-1}{12}  \mu$, $r_2=\frac{1}{6}\mu$ and $r_3=-\frac{1+\sqrt{21}}{12} \mu$. To illustrate, we show the profiles of the amplitudes in $U$ and $\hat{U}$  at $\mu=1$ in Fig.~\ref{Fig1}.

\item[(ii)]
If $ r_1 = \frac{1}{3}\mu$ ($\mu>0$) and  $C_1=0$, solving $f(r)=0$ gives rise to $r_1= \frac{1}{3}\mu $, $r_2 =-\frac{1}{6}\mu + \frac{1}{2} \sqrt{C_0+\mu ^2} $ and $r_3=-\frac{1}{6}\mu -\frac{1}{2} \sqrt{C_0+\mu ^2} $ with $-\mu^2< C_0 < 0$. In this case,  $U$ and $\hat{U}$ can be expressed in the $\dn$-function form
\begin{align}
\left\{
\begin{array}l
U_{\rm{ii}} = \sqrt{2}\,\alpha_1\,\dn(\alpha_1x, m_1) e^{i \mu t}, \\[1.5mm]
\hat{U}_{\rm{ii}} = \sqrt{2}\,\alpha_1\,\dn(\alpha_1x, m_1) e^{-i \mu t},
\end{array} \right.
\label{Sol1}
\end{align}
where  $\alpha_1$ and $m_1$ are given by
\begin{align}
 \alpha_1=\sqrt{\frac{\mu }{2 - m_1}},  \quad m_1= \frac{2\sqrt{C_0+\mu
   ^2}}{\sqrt{C_0+\mu ^2}+\mu } \quad (-\mu^2< C_0 < 0,\,\, \mu>0). \label{alpham1}
\end{align}

\item[(iii)]
If $r_2=\frac{1}{3}\mu$ ($ \mu\in \mathbb{R}$) and  $C_1=0$, solving $f(r)=0$ gives rise to $r_1 =-\frac{1}{6}\mu + \frac{1}{2} \sqrt{C_0+\mu ^2} $, $r_2= \frac{1}{3}\mu $ and $r_3=-\frac{1}{6}\mu -\frac{1}{2} \sqrt{C_0+\mu ^2} $ with $ C_0 > 0$. For such case, $U$ and $\hat{U}$ can be written in the $\cn$-function form
\begin{align}
\left\{
\begin{array}l
U_{\rm{iii}} =\alpha_2\sqrt{2 m_2}\, \cn(\alpha_2 x, m_2) e^{i \mu t}, \\[1.5mm]
\hat{U}_{\rm{iii}} = \alpha_2\sqrt{2 m_2}\, \cn(\alpha_2 x, m_2) e^{-i \mu t},
\end{array}
\right.  \label{Sol2}
\end{align}
where  $\alpha_2$ and $m_2$ are given by
\begin{align}
\alpha_2 = \sqrt[4]{C_0+\mu ^2}, \quad m_2= \frac{1}{2} + \frac{\mu }{2\sqrt{C_0+\mu
   ^2}},  \quad (C_0 > 0).  \label{alpham2}
\end{align}

\end{enumerate}

Second, we consider $\sigma=1, \gamma=0 $ and $W(x)\le 0$  for all $x\in \mathbb{R}$. In this case, there must be $W(0)=2\sigma (\frac{\mu}{3}-r_{3})=0$, which means $ r_3 = \frac{1}{3}\mu$. Meanwhile, $f(r_3) = f(\frac{1}{3}\mu)=\frac{C_1}{4} $ shows $C_1=0$. Correspondingly, the three roos of $f(r)=0$ are obtained as $r_1 =-\frac{1}{6}\mu + \frac{1}{2} \sqrt{C_0+\mu ^2} $, $r_2 =-\frac{1}{6}\mu -\frac{1}{2} \sqrt{C_0+\mu ^2} $ and $ r_3 = \frac{1}{3}\mu$ with $-\mu^2< C_0 < 0$. Thus, we obtain $U$ and $\hat{U}$  in the $\sn$-function form
\begin{align}
\left\{
\begin{array}l
U_{\rm{iv}} = \alpha_3 \sqrt{2 m_3}\,\sn(\alpha_3x, m_3) e^{i (\mu t+\frac{\pi}{2})}, \\[1.5mm]
\hat{U}_{\rm{iv}} = -\alpha_3 \sqrt{2 m_3}\,\sn(\alpha_3x, m_3) e^{-i (\mu t+\frac{\pi}{2})},
\end{array}
\right.  \label{Sol3}
\end{align}
where  $\alpha_3$ and $m_3$ are given by
\begin{align}
\alpha_3= \sqrt{\frac{-\mu }{1+m_3}}, \quad m_3 =\frac{\mu + \sqrt{C_0+\mu ^2}}{\mu -\sqrt{C_0+\mu ^2}},   \quad (-\mu^2< C_0 < 0,\,\, \mu<0).  \label{alpham3}
\end{align}

Third, when $\sigma=-1, \gamma=0$, it is impossible to get the smooth solutions of $U$ and $\hat{U}$. The reason can be seen as follows:  The necessary condition $W(0)\ge 0$ implies that $ r_3 \ge \frac{1}{3}\mu $ with $ \mu < 0$. But $f(r)< 0$ for $r\in(-\infty, r_3)$ and $f(\frac{1}{3}\mu)=\frac{C_1}{4}\ge 0$ shows that $C_1$ must be equal to $0$. Thus, the three real roots of $f(r)=0 $ are obtained as $r_1 =-\frac{1}{6}\mu + \frac{1}{2} \sqrt{\mu ^2-C_0} $,  $r_2=-\frac{1}{6}\mu -\frac{1}{2} \sqrt{\mu ^2-C_0} $ and $r_3= \frac{1}{3}\mu $ with $ 0< C_0 < \mu^2$. Then, substituting them into~\eref{Vsolution} gives
\begin{align}
W(x) =  -\big(\mu + \sqrt{\mu ^2-C_0} \big)\, \sn^2\bigg( \sqrt{\frac{\sqrt{\mu ^2-C_0}-\mu}{2}}\,x,  \frac{\mu +\sqrt{\mu ^2-C_0}}{\mu -\sqrt{\mu^2-C_0}}\bigg)\ge 0. \label{Vxexap}
\end{align}
However, this does not yield the smooth solutions $U$ and $\hat{U}$ which satisfy the symmetric relation $\hat{U}(x,t) = U(-x,t)$.

Next, by taking $\gamma=K$ in Eq.~\eref{Vsolution}, we consider all the smooth stationary solutions when $r_1 > r_2 >r_3$. One should note that the shift $x\to x+ K$ does not influence the sign definiteness of $W(x)$.  Therefore, with the same parametric conditions, we can obtain the other four Jacobi elliptic-function solutions:
\begin{align}
& \left\{
\begin{array}l
U^{(K)}_{\rm{i}}= \sqrt{\frac{2\mu}{3}-2r_{3}-2(r_{2}-r_{3})\sn^2(\sqrt{r_{1}-r_{3}}\,x+K, m)} \\[1mm]
\hspace{1.3cm}  \times  e^{-\frac{3 C_2}{4}\left[\frac{3 \left(r_2-r_1\right)}{\left(\mu -3 r_1\right) \left(\mu -3 r_2\right)\sqrt{r_1-r_3}} \Pi\left(\frac{m\left(\mu -3 r_1\right)}{\mu -3 r_2}; \phi(x), m\right)+ \frac{x}{\mu -3 r_1}\right] + i \, \mu t},  \\[2mm]
\hat{U}^{(K)}_{\rm{i}}=  \sqrt{\frac{2\mu}{3}-2r_{3}-2(r_{2}-r_{3})\sn^2(\sqrt{r_{1}-r_{3}}\,x + K, m)},  \\[1mm]
\hspace{1.3cm}  \times  e^{\frac{3 C_2}{4}\left[\frac{3 \left(r_2-r_1\right)}{\left(\mu -3 r_1\right) \left(\mu -3 r_2\right)\sqrt{r_1-r_3}} \Pi\left(\frac{m\left(\mu -3 r_1\right)}{\mu -3 r_2}; \phi(x), m\right)+ \frac{x}{\mu -3 r_1}\right] - i \, \mu t},
\end{array}\right. \,\, (\mu>0), \label{uxK} \\
& \left\{
\begin{array}l
U^{(K_1)}_{\rm{ii}} = \sqrt{2}\,\alpha_1\,\dn(\alpha_1x+K_1, m_1) e^{i \mu t}, \\[1.5mm]
\hat{U}^{(K_1)}_{\rm{ii}} = \sqrt{2}\,\alpha_1\,\dn(\alpha_1x+K_1, m_1) e^{-i \mu t},
\end{array}
\right.  \,\,(\mu>0),  \label{Sol1b}  \\
& \left\{
\begin{array}l
U^{(K_2)}_{\rm{iii}} =\alpha_2 \sqrt{2 m_2}\,\cn(\alpha_2 x+K_2, m_2) e^{i \mu t} , \\[1.5mm]
\hat{U}^{(K_2)}_{\rm{iii}} = -\alpha_2 \sqrt{2 m_2}\,\cn(\alpha_2 x+K_2, m_2) e^{-i \mu t},
\end{array}
\right.  \,\, (\mu\in \mathbb{R}),  \label{Sol2b} \\[2mm]
&
\left\{
\begin{array}l
U^{(K_3)}_{\rm{iv}} = \alpha_3 \sqrt{2 m_3}\,\sn(\alpha_3x+K_3, m_3) e^{i (\mu t + \frac{\pi}{2})}, \\[1.5mm]
\hat{U}^{(K_3)}_{\rm{iv}} = \alpha_3 \sqrt{2 m_3}\,\sn(\alpha_3x+K_3, m_3) e^{-i (\mu t + \frac{\pi}{2})},
\end{array}
\right.  \,\, (\mu<0), \label{Sol3b}
\end{align}
where $K_i=K(m_i)$ as defined in~\eref{ComEllInt}, $\alpha_i$'s and $m_i$'s ($1\leq i \leq 3$) are given by Eqs.~\eref{alpham1},~\eref{alpham2} and~\eref{alpham3}.  Based on the properties
\begin{align}
\begin{array}l
\phi(x) = -\phi(-x), \quad  \sn(x+K, m) = \sn(-x+K, m), \\
\dn(x+K, m) = \dn(-x+K, m), \quad  \cn(x+K, m) = -\cn(-x+K, m),
\end{array}
\end{align}
we know that solutions~\eref{uxK} and~\eref{Sol1b} still satisfy Eq.~\eref{NNLS} with $\sigma=1$, whereas solutions~\eref{Sol2b} and~\eref{Sol3b} solve Eq.~\eref{NNLS} with $\sigma=-1$.

For the above bounded cases, solutions~\eref{Sol1},~\eref{Sol2},~\eref{Sol1b} and~\eref{Sol3b} are even-symmetric with respect to $x$, while solutions~\eref{Sol3} and~\eref{Sol2b} are odd-symmetric about $x$. Moreover, we point out that the solutions of Eq.~\eref{NNLS} remain invariant only for some special shift in $x$, which is quite different from the NLS equation. In addition, it should be mentioned that the bounded Jacobi elliptic-function solutions~\eref{Sol1},~\eref{Sol2}, \eref{Sol3},~\eref{Sol1b}--\eref{Sol3b} coincide with those obtained in Ref.~\cite{Khare}.

\subsection{Hyperbolic-function solutions with $\delta=0$}
\label{sec3.2}

In this subsection, we study the hyperbolic-function solutions obtained from the Jacobi elliptic-function solutions in subsection~\ref{sec3.1} when particularly taking $r_{1}=r_{2}>r_{3}$. Since $W(x)$ in~\eref{Vsolution}  reduces to a constant or $0$ when $r_2=r_3$, we just consider the reduction of solutions~\eref{ux},~\eref{Sol1},~\eref{Sol2} and~\eref{Sol3} at $r_1=r_2$.

\begin{figure}[H]
\centering
\subfigure[]{\label{Fig2a}
\includegraphics[width=2.4in]{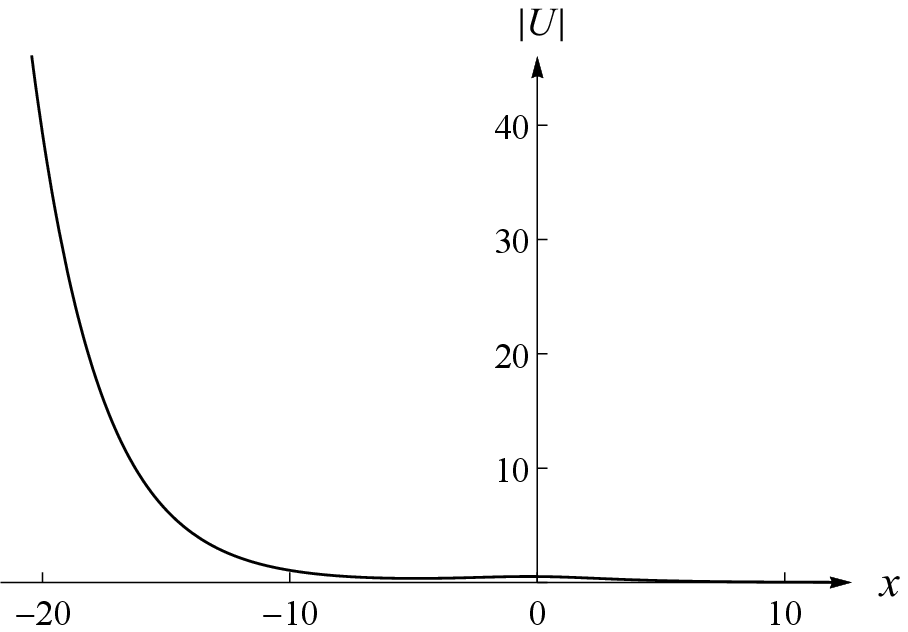}}\hspace{2.5cm}
\subfigure[]{ \label{Fig2b}
\includegraphics[width=2.4in]{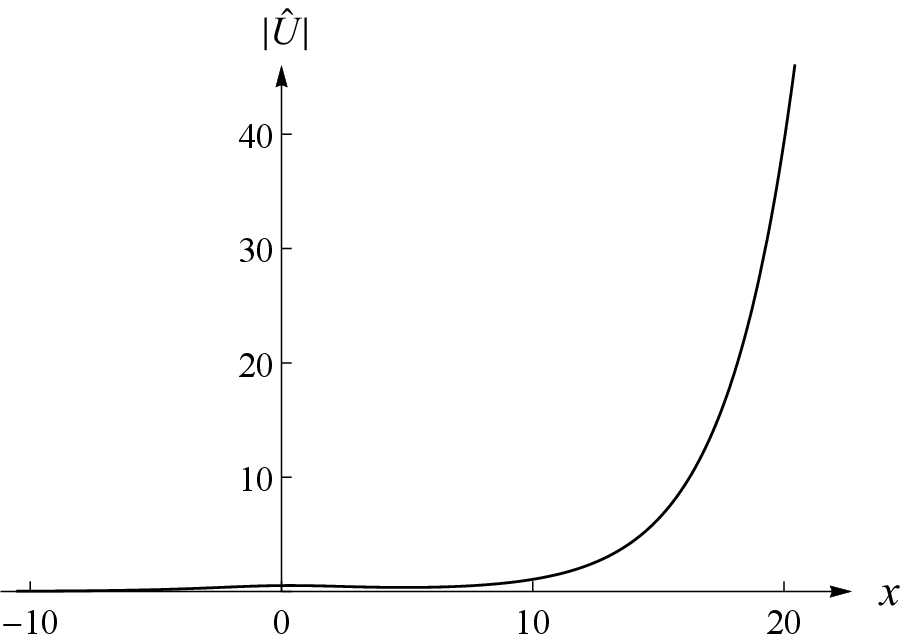}}
\caption{\small  Profiles of the amplitudes in $U$ and $\hat{U}$ via solution~\eref{ux2} with $\mu= \frac{4}{25} $, $r_1= \frac{1}{25}$ and $\epsilon = 1$. } \label{Fig2}
\end{figure}

When $\sigma=1$, if $0< r_1 =r_2 < \frac{1}{3}\mu$ ($\mu>0$) and $C_1>0$,  we have
$C_0= -\frac{4}{3} \left(\mu ^2-9 r_1^2\right)$ and $C_1= \frac{16}{27} \left(\mu -3 r_1\right){}^2 \left(\mu +6 r_1\right) $. Then, solution~\eref{ux} reduces to
\begin{align}
& \left\{
\begin{array}l
U_{\rm{v}} = \sqrt{\frac{2}{3}\,\mu +r_1\big[4- 6\tanh ^2(\sqrt{3r_1}\,x )\big]} e^{\epsilon\,\tanh ^{-1}
\left[3 \sqrt{\frac{r_1}{\mu +6 r_1}} \tanh \left(\sqrt{3r_1}\,x\right)\right]
-\epsilon\sqrt{\frac{\mu +6 r_1}{3}} x + i \mu t}, \\[2mm]
\hat{U}_{\rm{v}} = \sqrt{\frac{2}{3}\,\mu +r_1\big[4- 6\tanh ^2(\sqrt{3r_1}\,x )\big]} e^{\epsilon \sqrt{\frac{\mu +6 r_1}{3}} x
 -\epsilon\,\tanh ^{-1}
\left[3 \sqrt{\frac{r_1}{\mu +6 r_1}} \tanh \left(\sqrt{3r_1}\,x\right)\right]-i \mu t},
\end{array}\right.\,(\mu>0), \label{ux2}
\end{align}
with $\epsilon=\pm1,\, 0< r_1 < \frac{1}{3}\mu$. It can be seen that solution~\eref{ux2} is  non-symmetric with respect to $x$, and their amplitudes will grow to $\infty$ or decay to $0$ as $x \to \pm \infty$, as shown in Fig.~\ref{Fig2}.

If $ r_1 =r_2= \frac{1}{3}\mu$ ($\mu>0$), one naturally has  $C_0=C_1=0$. For such case, both solutions~\eref{Sol1} and~\eref{Sol2} become the same bright-soliton solution:
\begin{align}
\left\{
\begin{array}l
U_{\rm{vi}} = \sqrt{2\mu}\,\sech(\sqrt{\mu}\,x)e^{i \mu t}, \\[1.5mm]
\hat{U}_{\rm{vi}} = \sqrt{2\mu}\,\sech(\sqrt{\mu}\,x)e^{-i \mu t},
\end{array}
\right.  \,\,  (\mu>0) .  \label{Ssech1}
\end{align}
If $r_1=r_2$ and $ r_3 = \frac{1}{3}\mu\,\,(\mu<0)$, we have
$C_0=-\mu^2$ and $C_1=0$. Thus, solution~\eref{Sol3} becomes the dark-soliton solution:
\begin{align}
\left\{
\begin{array}l
U_{\rm{vii}} = \sqrt{-\mu}\, \tanh(\sqrt{-\mu/2}\,x)e^{i(\mu t + \frac{\pi}{2})},\\[1.5mm]
\hat{U}_{\rm{vii}} = -\sqrt{-\mu}\, \tanh(\sqrt{-\mu/2}\,x)e^{-i (\mu t + \frac{\pi}{2})},
\end{array}
\right.  \,\, (\mu<0). \label{Stanh1}
\end{align}
For $\sigma=-1$ and $r_1=r_2$, $W(x)$ in Eq.~\eref{Vxexap} reduces to the one given in Eq.~\eref{ExamVx}. As discussed in remark 1 below proposition~\ref{proposition1}, it is also impossible to derive the smooth solution in such particular case.

We notice that the stationary solutions~\eref{ux2}--\eref{Stanh1} solve Eq.~\eref{NNLS} with $\sigma=1$, but no hyperbolic-function solution is available for the defocusing NNLS equation. In addition, we mention that the stationary bright- and dark-soliton solutions~\eref{Ssech1} and~\eref{Stanh1} were first obtained in Ref.~\cite{Sarma}.

\subsection{Jacobi elliptic-function and hyperbolic-function solutions with $\delta \neq 0$}
\label{sec3.3}

In this part, we consider the general case $\delta \neq 0$ when $W(x)$ in~\eref{Vsolution} is substituted into Eqs.~\eref{ComplexVu11} and~\eref{StaSol}. In fact, Eq.~\eref{NNLS} is invariant with the imaginary translation transformation $U(x,t) \to U(x  + i \delta, t) $, which is a particular case of the Galilean-invariant transformation $U(x,t) \to U(x - 2 i v t + i \delta, t)e^{-v x+i v^2 t} \, (v,\delta\in\mathbb{R}) $~\footnote[4]{With the Galilean-invariant transformation, one can obtain more general solutions which may develop singularities at finite $t$~\cite{Ablowitz1,Ablowitz3}.}~\cite{Ablowitz3,Magstr}. Therefore, based on the results in subsections~\ref{sec3.1} and~\ref{sec3.2}, we can derive new complex-amplitude stationary solutions by directly making the replacement $x \to x+i\delta$. Meanwhile, we must consider the analyticity of the Jacobi elliptic functions and hyperbolic functions in the complex plane, and thus require the value of $\delta$ avoid the singularities of those functions. Here, one should be mentioned that $\sn(z)$, $\cn(z)$ and $\dn(z)$ are all doubly-periodic functions, while $\sech(z)$ and $\tanh(z)$ are both the periodic functions with the period $2\pi i$; and that $\sn(z)$ and $\cn(z)$ are both analytic except the points congruent to $iK'$ or $2K+iK'$, $\dn(z)$ is analytic except the points congruent to $iK'$ or $3iK'$, $\sech(z)$ and $\tanh(z)$ are both analytic except the points congruent to $ \frac{\pi}{2}i$ or $ \frac{3\pi}{2}i$~\cite{whittaker}.

First, from Eq.~\eref{ux},~\eref{uxK} and~\eref{ux2}, we obtain the following three unbounded Jacobi elliptic-function solutions with complex amplitudes for Eq.~\eref{NNLS} when $\sigma=1$:
\begin{align}
& \left\{
\begin{array}l
U^{(\delta)}_{\rm{i}}= \sqrt{\frac{2\mu}{3} - 2r_{3}-2(r_{2}-r_{3})\sn^2(\sqrt{r_{1}-r_{3}}\,x+ i \delta, m)}\,e^{\frac{- 3  C_2 \Pi \big(\frac{3 \left(r_2-r_3\right)}{\mu -3 r_3}; \phi_\delta(x), m\big)}{4 \left(\mu -3 r_3\right) \sqrt{r_1-r_3}} + i \, \mu t},  \\[2mm]
\hat{U}^{(\delta)}_{\rm{i}}= \sqrt{\frac{2\mu}{3} - 2r_{3}- 2(r_{2}-r_{3})\sn^2(\sqrt{r_{1}-r_{3}}\,x+ i \delta, m)}\,e^{\frac{ 3  C_2 \Pi \big(\frac{3 \left(r_2-r_3\right)}{\mu -3 r_3}; \phi_\delta(x), m\big) }{4\left(\mu -3 r_3\right)\sqrt{r_1-r_3}} - i \, \mu t},
\end{array}\right.  (\mu>0), \label{uxdel}  \\
& \left\{
\begin{array}l
U^{(K, \delta)}_{\rm{i}}= \sqrt{\frac{2\mu}{3}-2r_{3}-2(r_{2}-r_{3})\sn^2(\sqrt{r_{1}-r_{3}}\,x+ K + i \delta, m)} \\[1mm]
\hspace{2.1cm}  \times  e^{-\frac{3 C_2}{4}\left[\frac{3 \left(r_2-r_1\right)}{\sqrt{r_1-r_3}\left(\mu -3 r_1\right) \left(\mu -3 r_2\right)} \Pi\left(\frac{m\left(\mu -3 r_1\right)}{\mu -3 r_2}; \phi_\delta(x), m\right)+ \frac{\sqrt{r_1-r_3} x + i\delta}{\sqrt{r_1-r_3}(\mu -3 r_1)}\right] + i \, \mu t},  \\[2mm]
\hat{U}^{(K,  \delta)}_{\rm{i}}=  \sqrt{\frac{2\mu}{3}-2r_{3}-2(r_{2}-r_{3})\sn^2(\sqrt{r_{1}-r_{3}}\,x + K + i \delta, m)},  \\[1mm]
\hspace{2.1cm}  \times  e^{\frac{3 C_2}{4}\left[\frac{3 \left(r_2-r_1\right)}{\sqrt{r_1-r_3} \left(\mu -3 r_1\right) \left(\mu -3 r_2\right)} \Pi\left(\frac{m\left(\mu -3 r_1\right)}{\mu -3 r_2}; \phi_\delta(x), m\right)+ \frac{\sqrt{r_1-r_3} x + i\delta}{\sqrt{r_1-r_3} (\mu -3 r_1)}\right] - i \, \mu t},
\end{array}\right. \, (\mu>0), \label{uxKdel} \\
& \left\{
\begin{array}l
U^{(\delta)}_{\rm{v}} = \sqrt{\frac{2}{3}\,\mu +r_1\big[4- 6\tanh ^2(\sqrt{3r_1}\,x +i\delta)\big]} \\
\hspace{2.1cm}  \times e^{\epsilon\,\tanh ^{-1}
\left[3 \sqrt{\frac{r_1}{\mu +6 r_1}} \tanh \left(\sqrt{3r_1}\,x +i\delta \right)\right]
-\frac{\epsilon}{3}\sqrt{\frac{\mu +6 r_1}{r_1}}\left(\sqrt{3 r_1}x+i \delta \right) + i \mu t}, \\[2mm]
\hat{U}^{(\delta)}_{\rm{v}} = \sqrt{\frac{2}{3}\,\mu +r_1\big[4- 6\tanh ^2(\sqrt{3r_1}\,x +i\delta)\big]} \\
\hspace{2.1cm}  \times  e^{
\frac{\epsilon}{3}\sqrt{\frac{\mu +6 r_1}{r_1}}\left(\sqrt{3 r_1}x+i \delta \right)
-\epsilon\,\tanh ^{-1}\left[3 \sqrt{\frac{r_1}{\mu +6 r_1}} \tanh \left(\sqrt{3r_1}\,x +i\delta \right)\right]
-i \mu t},
\end{array}\right.\,(\mu>0), \label{ux2del}
\end{align}
where $m =\frac{r_{2}-r_{3}}{r_{1}-r_{3}}$, $\phi_\delta(x)=\text{am}\left(\sqrt{r_1-r_3}\,x+i\delta, m\right)$, $\epsilon=\pm1$ and $0< r_1 < \frac{1}{3}\mu$. In view of the analyticity of $\sn(z)$ and $\tanh(z)$, we require $\delta \neq (2n+1)K'$ ($n\in \mathbb{Z}$) for solutions~\eref{uxdel} and~\eref{uxKdel}, and $\delta \neq \frac{(2n+1)\pi}{2}$ ($n\in \mathbb{Z}$) for solution~\eref{ux2del}. Similarly, the amplitudes of solutions~\eref{uxdel}--\eref{ux2del} grow to $\infty$ as $x$ tends to one infinity but decays to $0$ as $x$ goes to the other infinity.

Second, from solutions~\eref{Sol1},~\eref{Sol2},~\eref{Sol3} and~\eref{Sol1b}--\eref{Sol3b}, we derive four bounded Jacobi elliptic-function solutions with complex amplitudes for Eq.~\eref{NNLS} when $\sigma=1$:
\begin{align}
&
\left\{
\begin{array}l
U^{(\delta)}_{\rm{ii}} = \sqrt{2}\,\alpha_1\,\dn(\alpha_1 x+i \delta, m_1) e^{i \mu t}, \\[1.5mm]
\hat{U}^{(\delta)}_{\rm{ii}} = \sqrt{2}\,\alpha_1\,\dn(\alpha_1 x+i \delta, m_1) e^{-i \mu t},
\end{array} \right. \,\,(\mu>0), \label{Sol4} \\[2mm]
& \left\{
\begin{array}l
U^{(\delta)}_{\rm{iii}} = \alpha_2 \sqrt{2 m_2}\,\cn(\alpha_2 x+i \delta, m_2) e^{i \mu t}, \\[1.5mm]
\hat{U}^{(\delta)}_{\rm{iii}} =  \alpha_2 \sqrt{2 m_2}\,\cn(\alpha_2 x+i \delta, m_2) e^{-i \mu t},
\end{array} \right.  \,\,(\mu\in \mathbb{R}), \label{Sol5} \\[2mm]
& \left\{
\begin{array}l
U^{(\delta)}_{\rm{iv}}= \alpha_3 \sqrt{2 m_3}\,\sn(\alpha_3 x+i \delta, m_3) e^{i (\mu t+\frac{\pi}{2})}, \\[1.5mm]
\hat{U}^{(\delta)}_{\rm{iv}} = -\alpha_3 \sqrt{2 m_3}\,\sn(\alpha_3 x+i \delta, m_3) e^{-i (\mu t+\frac{\pi}{2})},
\end{array} \right.  \,\,(\mu<0), \label{Sol6}  \\[2mm]
& \left\{
\begin{array}l
U^{(K_1, \delta)}_{\rm{ii}} = \sqrt{2}\,\alpha_1\,\dn(\alpha_1 x+K_1+i\delta, m_1) e^{i \mu t},  \\[1.5mm]
\hat{U}^{(K_1, \delta)}_{\rm{ii}} = \sqrt{2}\,\alpha_1\,\dn(\alpha_1 x+K_1+i\delta, m_1) e^{-i \mu t} ,
\end{array} \right. \,\,(\mu>0), \label{Sol1c}
\end{align}
and two bounded Jacobi elliptic-function solutions with complex amplitudes for Eq.~\eref{NNLS} when $\sigma=-1$:
\begin{align}
& \left\{
\begin{array}l
U^{(K_2, \delta)}_{\rm{iii}} = \alpha_2 \sqrt{2 m_2}\,\cn(\alpha_2 x +K_2+i \delta, m_2)e^{i \mu t},   \\[1.5mm]
\hat{U}^{(K_2, \delta)}_{\rm{iii}} = -\alpha_2 \sqrt{2 m_2}\, \cn(\alpha_2 x +K_2+i \delta, m_2) e^{-i \mu t} ,
\end{array} \right.  \,\,(\mu\in \mathbb{R}), \label{Sol2c} \\[2mm]
& \left\{
\begin{array}l
U^{(K_3, \delta)}_{\rm{iv}} = \alpha_3 \sqrt{2 m_3}\,\sn(\alpha_3 x+K_3+i \delta, m_3) e^{i (\mu t+\frac{\pi}{2})}, \\[1.5mm]
\hat{U}^{(K_3, \delta)}_{\rm{iv}} = \alpha_3 \sqrt{2 m_3}\,\sn(\alpha_3 x+K_3+i \delta, m_3) e^{-i (\mu t+\frac{\pi}{2})},
\end{array} \right. \,\, (\mu<0), \label{Sol3c}
\end{align}
where  $\alpha_i$'s and $m_i$'s ($1\leq i \leq 3$) are given by Eqs.~\eref{alpham1},~\eref{alpham2} and~\eref{alpham3},  $K_i$'s ($1\leq i \leq 3$) are three complete elliptic integrals, and $\delta \neq (2n+1)K'_i$ ($n\in \mathbb{Z}$) to assure the analyticity of the solutions. It can be found that solutions~\eref{Sol4}--\eref{Sol3c} are no longer even- or odd-symmetric with respect to $x$. Instead, solutions~\eref{Sol4}--\eref{Sol1c} are said to be \PT-symmetric because $U=$\PT$ U$, whereas solutions~\eref{Sol2c} and \eref{Sol3c} are anti-\PT-symmetric because $U=-$\PT$U$.


Third, based on solutions~\eref{Ssech1} and~\eref{Stanh1}, we get two general hyperbolic soliton solutions for Eq.~\eref{NNLS} with $\sigma=1$ as follows:
\begin{align}
&  \left\{
\begin{array}l
U^{(\delta)}_{\rm{vi}} = \sqrt{2\mu}\,\sech(\sqrt{\mu}\,x+i \delta) e^{i \mu t}, \\[1.5mm]
\hat{U}^{(\delta)}_{\rm{vi}} = \sqrt{2\mu}\,\sech(\sqrt{\mu}\,x+i \delta) e^{-i \mu t},
\end{array}
\right.  \,\,  (\mu>0),  \label{Ssech2} \\
&  \left\{
\begin{array}l
U^{(\delta)}_{\rm{vii}} = \sqrt{-\mu}\, \tanh(\sqrt{-\mu/2}\,x+i \delta)e^{i (\mu t + \frac{\pi}{2})},\\[1.5mm]
\hat{U}^{(\delta)}_{\rm{vii}} =-\sqrt{-\mu}\, \tanh(\sqrt{-\mu/2}\,x+i \delta)e^{-i (\mu t + \frac{\pi}{2})},
\end{array}
\right.  \,\, (\mu<0), \label{Stanh2}
\end{align}
where $\delta \neq \frac{(2n+1)\pi }{2}$ ($n\in \mathbb{Z}$) to assure the analyticity of the solutions.

\begin{figure}[H]
\centering
\subfigure[]{\label{Fig3a}
\includegraphics[width=2.5in]{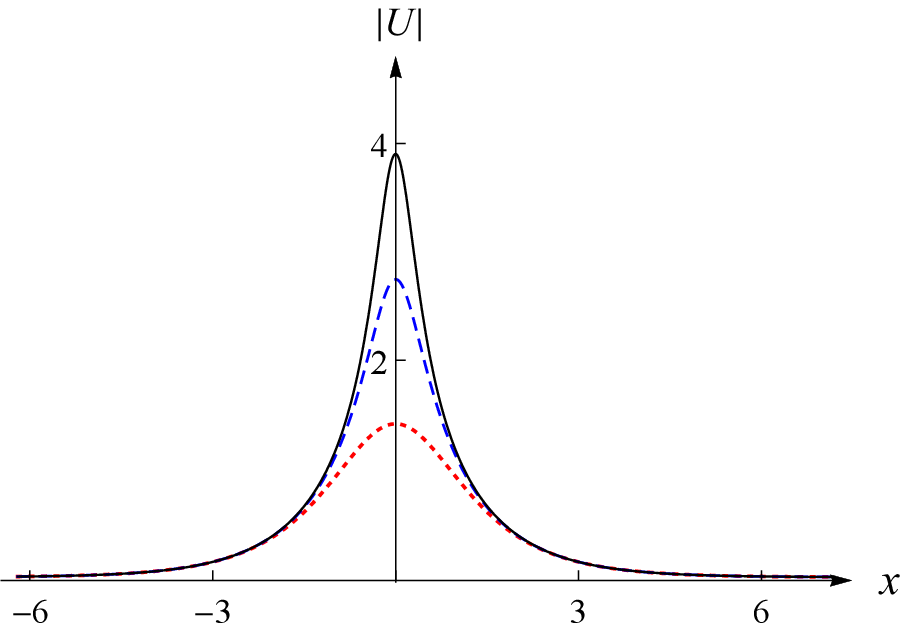}}\hspace{2.5cm}
\subfigure[]{ \label{Fig3b}
\includegraphics[width=2.5in]{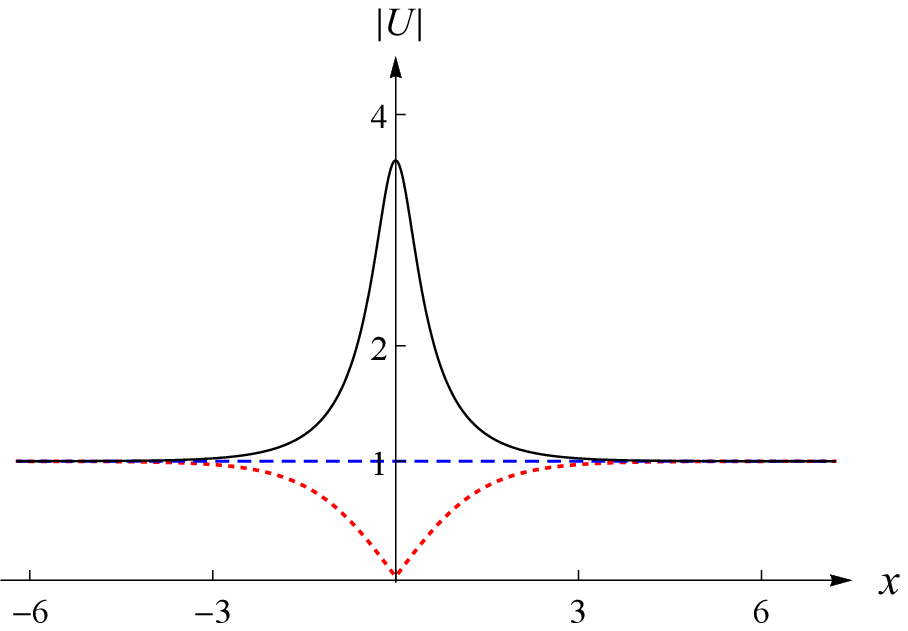}}
\caption{\small  (a) Soliton profiles via solution~\eref{Ssech2} at $\mu = 1$ with $\delta=0$ (red dotted),  $\delta=1$ (blue dashed), $\delta=\frac{6}{5} $ (black solid).  (b) Soliton profiles via solution~\eref{Stanh2} at $\mu = -1$ with $\delta=0$ (red dotted),  $\delta=\frac{\pi}{4} $ (blue dashed), $\delta=\frac{13}{10}$ (black solid).} \label{Fig3}
\end{figure}

In contrast to the common bright and dark solitons,  Eqs.~\eref{Ssech2} and~\eref{Stanh2} are both \PT-symmetric soliton solutions, and they can exhibit some unusual dynamical evolution behavior. As seen from Fig.~\ref{Fig3a}, solution~\eref{Ssech2} always displays the bright soliton profiles on the vanishing background.  The soliton amplitude and energy can be explicitly given by
\begin{align}
& A = \sqrt{2\mu}\,|\sec(\delta)| , \\
& E= \int_{-\infty}^{+\infty}|U^{(\delta)}_{\rm{vi}}|^2 d x =
\left\{
\begin{array}c
4 \sqrt{\mu }, \quad\quad\quad\, \delta = 0,   \\
 \frac{8 \sqrt{\mu}(\delta'-\frac{\pi}{2})}
{\sin\left(2 \delta \right)}, \quad \! \delta \neq 0,
\end{array} \right.
\end{align}
where $\delta'=\left(\delta +\frac{\pi }{2}\right) \bmod \pi$. For a given value of $\mu>0$, the amplitude and energy will increase with the increment of $\delta$ in the interval $ \big[n\pi , \frac{(2n+1)\pi }{2}\big)$ but decrease in the interval $\big(\frac{(2n+1)\pi }{2}, (n+1)\pi \big]$. Differently, solution~\eref{Stanh2} exists on the non-vanishing background and there is an asymptotic phase difference $\pi$ because $\lim_{x\ra \pm \infty} U^{(\delta)}_{\rm{vii}} =\pm \sqrt{-\mu} e^{i (\mu t + \frac{\pi}{2})}$.  Also, the soliton amplitude and energy  can be obtained by
\begin{align}
& A=\left\{
\begin{array}l
\sqrt{-\mu}\big(1 - | \tan \left(\delta\right)|\big), \quad
\delta\in \big[\frac{(4n-1)\pi}{4}, \frac{(4n+1)\pi}{4}\big], \\[2mm]
\sqrt{-\mu}\big( |\tan\left(\delta \right) | - 1 \big), \quad
\delta\in \big(\frac{(2n-1)\pi }{2}, \frac{(4n-1)\pi }{4}\big) \bigcup \big(\frac{(4n+1) \pi }{4}, \frac{(2n+1)\pi}{2}\big) ,
\end{array} \right. \label{Amp}  \\[2mm]
& E= \int_{-\infty}^{+\infty}\big||U^{(\delta)}_{\rm{vii}}|^2+\mu\big| d x =
\left\{
\begin{array}l
2 \sqrt{-2\mu }, \quad\quad\quad\, \delta = 0,   \\[1mm]
\frac{4\sqrt{-2\mu}\,|\delta'-\frac{\pi}{2}|}
{|\tan\left(2\delta \right)|}
, \quad \delta \neq 0,
\end{array} \right.
\end{align}
where $\delta'=\left(\delta +\frac{\pi }{2}\right) \bmod \pi$. From Eq.~\eref{Amp}, one can see that solution~\eref{Stanh2} represents the dark soliton for  $\delta\in \big(\frac{(4n-1)\pi}{4}, \frac{(4n+1)\pi}{4}\big)$ and anti-dark soliton for $\delta\in \big(\frac{(2n-1)\pi}{2}, \frac{(4n-1)\pi}{4}\big) \bigcup \big(\frac{(4n+1) \pi}{4}, \frac{(2n+1)\pi}{2}\big)$.  Particularly when $\delta= \frac{(4n+1) \pi}{4}$, solution~\eref{Stanh2} can be written as
\begin{align}
\left\{
\begin{array}l
U^{(\delta)}_{\rm{vii}} = -\sqrt{-\mu }e^{-i \tan ^{-1}\left[\sinh \left(\sqrt{-2\mu} x\right)\right] + i  \mu t }, \\[1.5mm]
\hat{U}^{(\delta)}_{\rm{vii}} = -\sqrt{-\mu }e^{-i \tan ^{-1}\left[\sinh \left(\sqrt{-2\mu} x\right)\right]-i  \mu t},
\end{array}\right.  \,\, (\mu<0). \label{Stanh3}
\end{align}
Interestingly, although the phase of such solution depends on $x$, its amplitude exhibits no spatial-localization in the $x$-coordinate. In Fig.~\ref{Fig3b}, we show the three different profiles displayed by solution~\eref{Stanh2}.

\section{Complex \PT-symmetric potentials}
\label{Sec4}

We recall that Eq.~\eref{NNLS} can be viewed as a linear Schr\"{o}dinger equation~\eref{PTLS} with the self-induced \PT-symmetric potential $V(x,t)=\sigma U(x,t)U^*(-x,t)$. Based on the stationary solutions in subsection~\ref{sec3.3}, we can obtain a class of complex and time-independent \PT-symmetric potentials:
\begin{align}
V(x; \delta) = 2\left[\frac{\mu}{3}-r_{3}-(r_{2}-r_{3})\sn^2\left(\tilde{x} + i \delta, m\right)\right],\quad \Big(m =\frac{r_{2}-r_{3}}{r_{1}-r_{3}} \Big),  \label{PTV}
\end{align}
where $\tilde{x}=\sqrt{r_1-r_3}\,x$ or $\sqrt{r_1-r_3}\,x +K $.  Meanwhile, the amplitude functions $u= U/e^{i \, \mu t} $ and $\hat{u}= \hat{U}/e^{-i \, \mu t} $ of  those solutions may correspond to some eigenfunctions of the stationary Schr\"{o}dinger equation
\begin{align}
H_\delta \psi = \lambda \psi, \quad H_\delta:= \frac{d^2}{dx^2} + V(x; \delta),  \label{PTLS2}
\end{align}
at the eigenvalue $\lambda=\mu$.
It should be noted that  the amplitudes of solutions~\eref{uxdel}--\eref{ux2del} are not associated to any eigenfunction of Eq.~\eref{PTLS2} since they grow to infinite as $x\to \infty$ or $ -\infty$, which violates the square-integrability on $\mathbb{R}$.


Here, we present the complex \PT-symmetric potentials associated with the bounded solutions~\eref{Sol4}--\eref{Stanh2} as follows:

(a) $\sigma=1$, $r_1= \frac{1}{3}\mu, r_2 =-\frac{1}{6}\mu + \frac{1}{2} \sqrt{C_0+\mu ^2}, r_3=-\frac{1}{6}\mu -\frac{1}{2} \sqrt{C_0+\mu ^2} $ ($\mu>0,\,-\mu^2< C_0 < 0$) and $C_1=0$.
\begin{align}
V_{\rm{ii}}(x; \delta) = & \big(\mu + \sqrt{C_0+\mu ^2} \big) \left\{ \frac{
   \text{cn}^2(\delta, m'_1) \text{dn}^2(\delta, m'_1) \text{dn}^2(\tilde{x}, m_1) - m_1^2\text{sn}^2(\delta, m'_1)\text{cn}^2(\tilde{x}, m_1) \text{sn}^2(\tilde{x}, m_1)}
{\big[\text{cn}^2(\delta, m'_1) + m_1 \text{sn}^2(\delta, m'_1) \text{sn}^2(\tilde{x}, m_1)\big]^2} \right.  \notag \\
& \hspace{2.8cm}  \left. -\, i \frac{2\,m_1 \text{dn}(\delta, m'_1)\text{cn}(\delta, m'_1)
\text{sn}(\delta, m'_1) \text{cn}(\tilde{x}, m_1)\text{dn}(\tilde{x}, m_1) \text{sn}(\tilde{x}, m_1)}
{\big[\text{cn}^2(\delta, m'_1) + m_1 \text{sn}^2(\delta, m'_1) \text{sn}^2(\tilde{x}, m_1)\big]^2} \right\},  \label{PTii}
\end{align}
where $\tilde{x}=\alpha_1 x$ or $\alpha_1 x+ K_1$, and $m'_1=1-m_1$ with $\alpha_1$ and $m_1$ defined
by~\eref{alpham1}.

(b) $\sigma=\pm 1$, $r_1 =-\frac{1}{6}\mu + \frac{1}{2} \sqrt{C_0+\mu ^2}, r_2= \frac{1}{3}\mu, r_3=-\frac{1}{6}\mu -\frac{1}{2} \sqrt{C_0+\mu ^2} $ ($ \mu\in \mathbb{R}, C_0 > 0$) and $C_1=0$.
\begin{align}
V_{\rm{iii}}(x; \delta) = &
\big(\mu + \sqrt{C_0+\mu ^2}\big) \left\{ \frac{
\text{cn}^2(\delta, m'_2)\text{cn}^2(\tilde{x}, m_2) -\text{dn}^2 (\delta, m'_2)
\text{sn}^2(\delta, m'_2)\text{dn}^2(\tilde{x}, m_2)
\text{sn}^2(\tilde{x}, m_2)}{\big[\text{cn}^2(\delta, m'_2) + m_2 \text{sn}^2(\delta, m'_2) \text{sn}^2(\tilde{x}, m_2)\big]^2} \right.  \notag \\
& \hspace{2.8cm}  \left. -\,i \frac{2\,\text{dn}(\delta, m'_2)
\text{cn}(\delta, m'_2) \text{sn}(\delta, m'_2)
\text{cn}(\tilde{x}, m_2)\text{dn}(\tilde{x}, m_2)\text{sn}(\tilde{x}, m_2)}
{\big[\text{cn}^2(\delta, m'_2) + m_2 \text{sn}^2(\delta, m'_2) \text{sn}^2(\tilde{x}, m_2)\big]^2} \right\},   \label{PTiii}
\end{align}
where $\tilde{x}=\alpha_2 x$ for $\sigma=1$ and $\tilde{x} = \alpha_2 x+ K_2$ for $\sigma=-1$, and $m'_2=1-m_2$ with
$\alpha_2$ and $m_2$ defined by~\eref{alpham2}.

(c) $\sigma=\pm 1$, $r_1 =-\frac{1}{6}\mu + \frac{1}{2} \sqrt{C_0+\mu ^2}, r_2 =-\frac{1}{6}\mu -\frac{1}{2} \sqrt{C_0+\mu ^2},  r_3 = \frac{1}{3}\mu$ ($\mu<0, -\mu^2< C_0 < 0$) and $C_1=0$.
\begin{align}
V_{\rm{iv}}(x; \delta) = & \big(\mu+ \sqrt{C_0+\mu ^2} \big)\left\{
\frac{\text{dn}^2(\delta, m'_3)\text{sn}^2(\tilde{x}, m_3)
-\text{sn}^2(\delta, m'_3)\text{cn}^2(\delta, m'_3) \text{cn}^2(\tilde{x}, m_3) \text{dn}^2(\tilde{x}, m_3)}
{\big[\text{cn}^2(\delta, m'_3) + m_3 \text{sn}^2(\delta, m'_3) \text{sn}^2(\tilde{x}, m_3)\big]^2} \notag \right. \notag \\
& \hspace{2.8cm}  \left. +\,i \frac{2\, \text{dn}(\delta, m'_3)
\text{cn}(\delta, m'_3) \text{sn}(\delta, m'_3)
\text{cn}(\tilde{x}, m_3)\text{dn}(\tilde{x}, m_3) \text{sn}(\tilde{x}, m_3)}
{\big[\text{cn}^2(\delta, m'_3) + m_3 \text{sn}^2(\delta, m'_3) \text{sn}^2(\tilde{x}, m_3)\big]^2}
\right\},   \label{PTiv}
\end{align}
with $\tilde{x}=\alpha_3 x$ for $\sigma=1$ and $\tilde{x}=\alpha_3 x+ K_3$  for $\sigma=-1$, and $m'_2=1-m_2$ with
$\alpha_3$ and $m_3$ defined by~\eref{alpham3}.

(d) $\sigma=1$,  $ r_1 =r_2= \frac{1}{3}\mu, r_3= -\frac{2}{3}\mu$ ($\mu>0$) and $C_0=C_1=0$.
\begin{align}
& V_{\rm{vi}}(x; \delta) = \frac{8 \mu\left[\cos^2(\delta)
\cosh^2(\sqrt{\mu } \,x )-\sin^2(\delta)\sinh^2 (\sqrt{\mu }\,x)\right]}
{\left[\cos (2 \delta)+\cosh(2 \sqrt{\mu } \,x )\right]^2}-\, i\frac{4 \mu  \sin (2\delta)\sinh
(2 \sqrt{\mu } \,x)}{\left[\cos (2 \delta) +\cosh (2 \sqrt{\mu} \,x)\right]^2}.  \label{PTvi}
\end{align}

(e) $\sigma=1$, $ r_1 =r_2= -\frac{1}{6}\mu, r_3 = \frac{1}{3}\mu\,\,(\mu<0)$, $C_0=-\mu^2$ and $C_1=0$.
\begin{align}
V_{\rm{vii}}(x; \delta) = \frac{\mu\left[\sinh^2\!\left(\sqrt{-2\mu}\,x\right)
-\sin^2\!\left(2\delta\right)\right]}
{\left[\cos \left(2\delta\right)
+\cosh \left(\sqrt{-2\mu}\,x\right)\right]^2} +\,i \frac{2\mu
\sin \left(2\delta\right)
\sinh \left(\sqrt{-2\mu } \,x\right)}
{\left[\cos\left(2\delta\right)+\cosh\left(\sqrt{-2\mu}\,x\right)\right]^2}.  \label{PTvii}
\end{align}

\begin{figure}[H]
\centering
\subfigure[]{\label{Fig4a}
\includegraphics[width=2.5in]{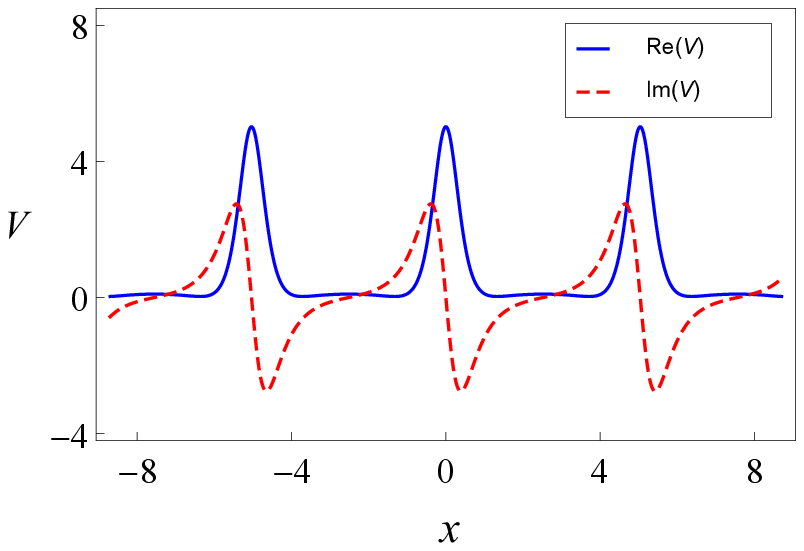}}\hspace{2.5cm}
\subfigure[]{ \label{Fig4b}
\includegraphics[width=2.5in]{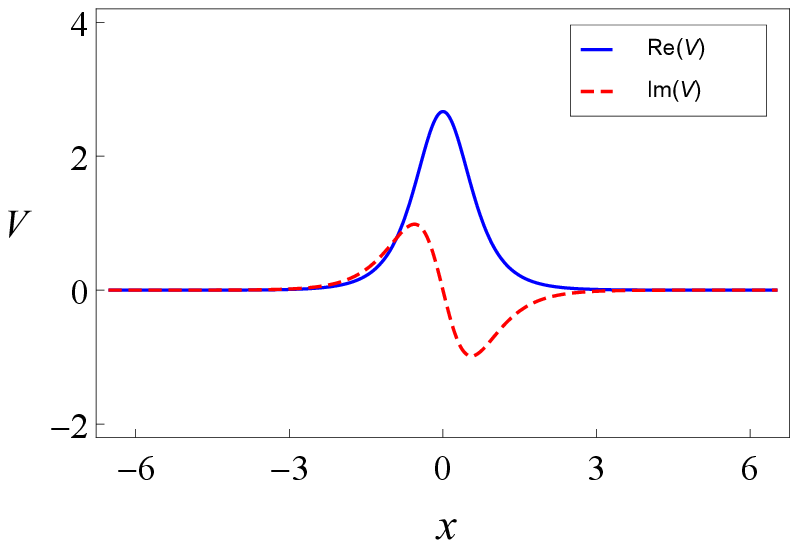}}
\caption{\small  (a) Profile of the Jacobi periodic \PT-symmetric potential via Eq.~\eref{PTii} with
$C_0=-0.5$, $\mu=1$ and $\delta=1.1$. (b) Profile of the hyperbolic localized \PT-symmetric potential via Eq.~\eref{PTvi} with
 $\mu=1$ and $\delta=\frac{\pi}{6}$. } \label{Fig4}
\end{figure}

For the above \PT-symmetric potentials, we require  $\delta \neq (2n+1)K'_i$ ($1\leq i\leq 3, n\in \mathbb{Z}$) in~\eref{PTii}--\eref{PTiv}, and $\delta \neq \frac{(2n+1)\pi}{2}$ ($n\in \mathbb{Z}$) in~\eref{PTvi} and~\eref{PTvii} to assure the analyticity; and require $\delta \neq n K'_i$ ($1\leq i\leq 3, n\in \mathbb{Z}$) in~\eref{PTii}--\eref{PTiv}, and $\delta \neq \frac{n \pi}{2}$ ($n\in \mathbb{Z}$) in~\eref{PTvi} and~\eref{PTvii} to avoid the vanishment of the imaginary parts.  With such nonsingular and nondegenerate conditions, Eqs.~\eref{PTii}--\eref{PTiv} are bounded, Jacobi periodic \PT-symmetric potentials (e.g., see Fig.~\ref{Fig4a}), whereas Eqs.~\eref{PTvi} and~\eref{PTvii} are bounded, hyperbolic localized \PT-symmetric potentials (e.g., see Fig.~\ref{Fig4b}). Correspondingly, the amplitudes of solutions~\eref{Sol4}-\eref{Sol3c} are the  eigenfunctions of Eq.~\eref{PTLS2} with the periodic conditions $\psi(x)=\psi(x+L)$ (where $L=\frac{2K_1}{\alpha_1}$ for~\eref{Sol4} and~\eref{Sol1c}, $L=\frac{4K_2}{\alpha_2}$ for~\eref{Sol5} and~\eref{Sol2c}, and $L=\frac{4K_3}{\alpha_3}$ for~\eref{Sol6} and~\eref{Sol3c}), the amplitude of solution~\eref{Ssech2} is an eigenfunction with the zero boundary condition $\psi(\pm \infty) =0,\, \psi_x(\pm \infty) =0$, and the amplitude of solution~\eref{Stanh2} is an eigenfunction with the nonzero boundary condition $\psi(\pm \infty) =\text{const.},\, \psi_x(\pm \infty) =0$.

Usually, the PTLS equation admits two  parametric regions~\cite{bender1}: (i) In the unbroken region of \PT~symmetry, all of the eigenvalues are  real, and $\psi$ is simultaneously an eigenfunction of the Hamiltonian $H$ and the combined operator \PT. (ii) In the broken region of \PT~symmetry,  there are a finite number of real and infinite number of complex conjugate pairs of eigenvalues, and some of the eigenfunctions of $H$ are not simultaneously eigenfunctions of \PT.  Thus, the \PT-symmetric system can exhibit a spontaneous symmetry breaking from the unbroken to broken \PT~phase when the non-Hermiticity parameter
exceeds a certain critical value.  However, we find that the symmetry breaking does not occur in Eq.~\eref{PTLS2} for the complex potentials in Eqs.~\eref{PTii}--\eref{PTvii} although they contain an arbitrary constant $\delta$.

\begin{proposition} \label{proposition3}
The Hamiltonian $H_\delta$ given in Eq.~\eref{PTLS2}  admits the all-real spectrum for all cases in Eqs.~\eref{PTii}--\eref{PTvii}, i.e., its \PT~symmetry remains unbroken.
\end{proposition}

\beginproof Assume that $\psi(x)$ is an eigenfunction of Eq.~\eref{PTLS2} at the eigenvalue $\lam$. With the transformation $x=z- i\delta $, Eq.~\eref{PTLS2} can be equivalently written as
\begin{align}
\frac{d^2}{dz^2}\varphi(z)  + V(z; 0)\varphi(z) = \lam \varphi(z), \label{PTLS3}
\end{align}
where $\varphi(z) := \psi(z-i\delta)$. Then, we continue the eigenvalue problem~\eref{PTLS3} into the complex-$z$ plane. By using the analyticity of the Jacobi elliptic functions and hyperbolic functions and noticing that $\delta \neq K'_i, \frac{\pi}{2}$ ($1\leq i\leq 3$), we know that $V(z; 0)$ must be analytic in an infinite strip domain $D:=\{z|-\epsilon < \Im(z)< \epsilon\}$, so that $\varphi(z)$ is also an analytic solution in $D$. Since $D$ covers the whole real axis $\mathbb{R}$, then $\varphi(x)$ satisfies the eigenvalue problem
\begin{align}
H_0 \varphi = \lambda \varphi, \quad H_0:= \frac{d^2}{dx^2} + V(x; 0),  \label{PTLS4}
\end{align}
where $V(x; 0)$ is a real, bounded, even-symmetric potential. Conversely, for any eigenfunction $\varphi(x)$ of Eq.~\eref{PTLS4} at the eigenvalue $\lam$, we obtain that $\psi(x)=\varphi(x+i\delta)$ also satisfies Eq.~\eref{PTLS2}. Hence, the eigenvalue problems~\eref{PTLS2} and~\eref{PTLS3} enjoy the same set of eigenvalues. In view of the Hermiticity of $H_0$, we immediately know that $H_\delta$ possesses the all-real spectrum for any nonzero $\delta \in \mathbb{R}$.
\hfill \endproof

Taking Eq.~\eref{PTvi} with $\delta=-\frac{(2n+1)\pi}{4}$ ($n\in \mathbb{Z}$) as an example,  we can express the potential $V_{\rm{vi}}$ as
\begin{align}
V_{\rm{vi}} = 4 \mu  \text{sech}^2\left(2 \sqrt{\mu } \,x\right) + 4\,i(-1)^n \mu \tanh \left(2 \sqrt{\mu } \,x\right)
\text{sech}\left(2 \sqrt{\mu } \,x\right), \quad (\mu>0).  \label{Scarf}
\end{align}
This is a particular type of Scarf II potential which has a fixed ratio of the imaginary part to the real one. As discussed in Ref.~\cite{Ahmed}, the Hamiltonian $H$ associated to Eq.~\eref{Scarf} must be in the \PT-symmetry unbroken phase because $4\,(-1)^n \mu < 4 \mu +\frac{1}{4}$ for any $\mu > 0$.

\section{Conclusions and discussions}
\label{Sec5}

By establishing the relationship between Eq.~\eref{NNLS} and the elliptic equation~\eref{Veq}, we have obtained the general stationary solutions for the NNLS equation. Then, by considering the dependence of the smoothness on the involved constants in Eq.~\eref{Veq}, we have discussed all smooth stationary solutions in terms of the Jacobi elliptic functions and hyperbolic functions. When $C_1>0$ and $ r_3 < r_2 < r_1 <\frac{1}{3}\mu $, the unbounded  stationary solutions~\eref{ux} and~\eref{ux2}, which can exhibit an infinite growth behavior as $x\to \infty$ or $x\to -\infty$, are possessed only by the focusing NNLS equation. Particularly for $C_1=0$, the five types of bounded stationary solutions, which include the Jacobi elliptic-function solutions~\eref{Sol1},~\eref{Sol2} and~\eref{Sol3} as well as the hyperbolic soliton solutions~\eref{Ssech1} and~\eref{Stanh1}, are shared by the NLS and NNLS equations in common. In addition, the $K$-shifted Jacobi elliptic-function solutions~\eref{uxK}--\eref{Sol3b} also solve Eq.~\eref{NNLS} with $\sigma=1$ or $\sigma=-1$, but they do not yield new soliton solutions at $r_1 = r_2$. It should be pointed out that the unbounded solutions (see Eqs.~\eref{ux},~\eref{uxK} and~\eref{ux2}) with $C_1>0$ are reported here for the first time, whereas all the bounded solutions (see Eqs.~\eref{Sol1},~\eref{Sol2},~\eref{Sol3},~\eref{Sol1b}--\eref{Sol3b},~\eref{Ssech1} and~\eref{Stanh1}) with $C_1=0$ coincide with the results in Refs.~\cite{Sarma,Khare}.

Moreover, based on the imaginary translation invariance of the NNLS equation, we have obtained the complex-amplitude stationary solutions and have given their associated nonsingular conditions. For the bounded cases, solutions~\eref{Sol4}--\eref{Sol1c},~\eref{Ssech2} and~\eref{Stanh2} are \PT-symmetric, while solutions~\eref{Sol2c} and~\eref{Sol3c} are anti-\PT-symmetric. Of special interest, solution~\eref{Stanh2} can exhibit no spatial localization in addition to the dark and anti-dark soliton profiles, which is sharp contrast with the common dark soliton. From the viewpoint of physical applications, the complex-amplitude stationary solutions can be used to construct a wide class of complex and time-independent \PT-symmetric potentials, and their associated Hamiltonians have been proved to be \PT-symmetry unbroken. Particularly for the bounded cases, the amplitudes of solutions~\eref{Sol4}--\eref{Stanh2} correspond to the eigenstates of the stationary Schr\"{o}dinger equation~\eref{PTLS2} with the associated \PT-symmetric potentials.  Also, it implies that the PTLS equation may support both the exact analytical \PT-symmetric and anti-\PT-symmetric eigenstates with the periodic boundary condition.


It might be an interesting issue to study the stability of the bounded complex-amplitude solutions~\eref{Sol4}--\eref{Stanh2}, most of which have not been reported before.
Here, we examine their linear stability by perturbing those solutions in the form~\cite{Yang,XuPel}
\begin{equation}
U(x,t)=\big\{u_\delta(x)+[v(x)+w(x)]e^{\lambda t}+[\hat{v}(x)-\hat{w}(x)]e^{\lambda^{*}t}\big\}e^{i \,\mu t},  \label{PSS}
\end{equation}
where $u_\delta(x)$ represents the amplitude of solutions~\eref{Sol4}--\eref{Stanh2} (e.g., $u_\delta(x)=\sqrt{2\mu}\,\sech(\sqrt{\mu}\,x+i \delta)$ for solution~\eref{Ssech2}),
$v(x),w(x)\ll 1$ are normal-mode perturbations, and $\lambda$ is the eigenvalue of this normal mode. Inserting the perturbed solution~\eref{PSS} into Eq.~\eref{NNLS} and linearizing the resulting equation, we obtain the  linear-stability eigenvalue problem:
\begin{equation}
\mathbf{L}_\delta \mathbf{\Psi}=\lambda\mathbf{\Psi}, \quad \mathbf{\Psi}
=
\begin{pmatrix}
v \\
w
\end{pmatrix}, \label{NNLS_StabMP}
\end{equation}
with
\begin{align}
\mathbf{L}_\delta = i \,\begin{pmatrix}
0 & \mathcal{L}- \frac{\sigma}{2} (u^2_\delta + \hat{u}^2_\delta) \\
\mathcal{L} + \frac{\sigma}{2} (u^2_\delta + \hat{u}^2_\delta) & 0
\end{pmatrix},  \quad \mathcal{L}:= \frac{d^2}{dx^2}-\mu + 2 \sigma u_\delta \hat{u}_\delta, \label{NNLS_StabL}
\end{align}
where the diagonal elements are zeros because $u^2_\delta = \hat{u}^2_\delta$ holds for solutions~\eref{Sol4}--\eref{Stanh2}.
Particularly for $\delta=0$, Eq.~\eref{NNLS_StabMP} coincides with the linear-stability eigenvalue problem for the corresponding stationary solutions of the NLS equation~\eref{NLS}. On the other hand, it can be readily shown that  $\mathbf{L}_\delta \mathbf{\Psi}=\lambda\mathbf{\Psi}$ and $\mathbf{L}_0 \mathbf{\Psi}=\lambda\mathbf{\Psi}$ share the same linear-stability spectrum by the way of proving proposition~\ref{proposition3}. Therefore, solutions~\eref{Sol4}--\eref{Stanh2} enjoy the same linear-stability criteria as their counterparts of Eq.~\eref{NLS} do, respectively~\cite{VK,Barashenkov,Kartashov}. However, it does not say that those stationary solutions have the same orbital stability or instability in
Eqs.~\eref{NNLS} and~\eref{NLS}, as seen in Ref.~\cite{Genoud}.


\section*{Acknowledgments}
We would like to thank the anonymous referees for their valuable comments, and the helpful discussions with Professor Yehui Huang from North China Electric Power University. T. X. appreciates the hospitality of the Department of Mathematics \& Statistics at McMaster University during his visit in 2019.  This work was partially supported by the National Natural Science Foundation of China (Grant Nos. 11705284 and 61505054), by the Fundamental Research Funds of the Central Universities (Grant No. 2017MS051), and by the program of China Scholarship Council (Grant No. 201806445009).

\end{document}